\documentclass[fleqn,10pt]{wlscirep}
\usepackage[utf8]{inputenc}
\usepackage[T1]{fontenc}
\usepackage{subcaption}
\usepackage{comment}
\usepackage{array}
\usepackage{multirow}
\usepackage{graphicx}
\usepackage{ulem}
\usepackage{lineno}

\title{Microcoulomb-level electron beam and multi-Joule hard X-rays driven by a high-efficiency laser-plasma accelerator}

\author[1,2,*]{B.~Mahieu}
\author[1,3]{L.~Ribotte}
\author[1]{W.~Cayzac}
\author[1,2]{G.~Boutoux}
\author[4]{R.~Parreault}
\author[4]{J.~Gastineau}
\author[4]{E.~Lamoine}
\author[4]{F.~Audo}
\author[5,6]{R.~Babjak}
\author[3]{D.~Batani}
\author[4]{N.~Blanchot}
\author[1]{J.~L.~Bourgade}
\author[4]{M.~Brochier}
\author[4]{T.~Caillaud}
\author[1,7]{P.~Canel}
\author[4]{S.~Cavaro}
\author[4]{C.~Chappuis}
\author[4]{S.~Debesset}
\author[4]{R.~Diaz}
\author[3]{E.~D'Humières}
\author[1]{W.~Duchastenier}
\author[4]{R.~du Jeu}
\author[1]{A.~Duval}
\author[4]{B.~Etchessahar}
\author[4]{M.~Ferri}
\author[1]{M.~Garandeau}
\author[1,2]{L.~Gremillet}
\author[1]{V.~Hénot}
\author[4]{E.~Journot}
\author[8]{J.~C.~Kieffer}
\author[1]{I.~Lantuéjoul}
\author[4]{L.~Le Déroff}
\author[9]{N.~Lemos}
\author[1,2]{C.~Rousseaux}
\author[4]{F.~Scol}
\author[7]{K.~Ta Phuoc}
\author[4]{W.~Vaillant}
\author[1]{B.~Vauzour}
\author[5]{M.~Vranic}
\author[1,2,*]{X.~Davoine}
\author[9]{F.~Albert}

\affil[1]{CEA, DAM, DIF, F-91297 Arpajon, France}
\affil[2]{Universit\'e Paris-Saclay, CEA, LMCE, F-91680 Bruy\`eres-le-Ch\^atel, France}
\affil[3]{CELIA, Universit\'e de Bordeaux–CNRS–CEA, UMR 5107, 33405 Talence, France}
\affil[4]{CEA-CESTA, F-33116 Le Barp, France}
\affil[5]{GoLP/Instituto de Plasmas e Fusão Nuclear, Instituto Superior Técnico, Universidade de Lisboa, Lisbon, 1049-001, Portugal}
\affil[6]{Institute of Plasma Physics, Czech Academy of Sciences, U Slovanky 2525/1a, 182 00 Praha 8, Czechia}
\affil[7]{LOA, ENSTA Paris, CNRS, Ecole Polytechnique, Institut Polytechnique de Paris, 91762 Palaiseau, France}
\affil[8]{INRS-EMT, Universit\'e du Qu\'ebec, Varennes J3X 1S2, Qu\'ebec, Canada}
\affil[9]{Lawrence Livermore National Laboratory, Livermore, California 94550, USA}

\affil[*]{email: benoit.mahieu@cea.fr; xavier.davoine@cea.fr}

\begin{abstract}
We report on the production of ultrahigh-charge relativistic electron beams and the development of a laser-wakefield acceleration platform at the LMJ facility. Making use of the kilojoule-class, sub-picosecond PETAL laser pulse focused onto a supersonic helium gas jet, electron beams carrying a total charge beyond 1~µC were generated, with energies up to $\sim$500\,MeV.
Given the ps-scale laser pulse duration, an on-target intensity approaching $10^{19}~\mathrm{W/cm^2}$, and a plasma density reaching 2\% of the critical density, electron energisation arises from a combination of self-modulated laser wakefield acceleration (SMLWFA) and direct laser acceleration (DLA). The resulting electron spectrum exhibits a Maxwellian-like distribution, characteristic of this mixed SMLWFA/DLA regime.
The total energy carried by the electron beam is estimated to be up to 17~J, within a sub-ps duration.
A broadband Joule-level photon beam was also produced by Bremsstrahlung, demonstrating the potential for future applications. 
Experimental results are supported by start-to-end numerical simulations, including 3-D particle-in-cell and Monte-Carlo particle transport calculations. These findings pave the way for applications requiring high-charge electron beams, including the generation of high-power secondary radiation or particle sources. The use of these beams to probe matter in high-energy density states driven by the nanosecond-duration LMJ beams represents another promising avenue.

\end{abstract}
\begin{document}

\flushbottom
\maketitle
\thispagestyle{empty}

\section*{Introduction}

The production of high-charge relativistic electron beams via laser-wakefield acceleration (LWFA) has been increasingly explored in recent years. These beams serve as ideal drivers for bright, sub-picosecond secondary sources, such as hard X/$\gamma$-ray photons~\cite{Albert_PRL_2017, Albert_POP_2018, Lemos_PPCF_2018, Lemos2019, Li_HEDP_2020, Sinclair_IEEE_AAC_2022, Lemos_PRR_2024}, fast neutrons~\cite{Jiao_MRE_2017, Feng_HEDP_2020, Gunther_NatComm_2022}, and relativistic electron–positron pairs~\cite{Sarri_PRL_2013, Sarri_NatComm_2015, Alejo_PPCF_2020, Streeter_SciRep_2024}. In turn, these sources can support diverse applications, including ultrafast radiography~\cite{Lemos_PRR_2024, glinec2005high, Ramanathan2010,williams2017positron}, high-energy-density (HED) diagnostics~\cite{Mahieu_NatCom_2018}, nuclear physics~\cite{CHEN2024}, and laboratory astrophysics~\cite{Chen2023, Sarri_NatComm_2015}.

Many of these secondary processes do not require high beam quality or narrow energy spreads, but instead scale directly with the total flux of incident electrons. Consequently, generating primary electron beams at moderate energies (a few tens of MeV) is often advantageous. This applies, for instance, to generating electron–positron pair plasmas via the Bethe-Heitler process to mimic astrophysical phenomena~\cite{Chen2023}.
As another example, generating X-ray backlighters via inverse Compton scattering requires moderate-energy electron beams to produce photons in the relevant energy ranges: a few keV for matching material absorption edges in spectroscopy, or 10--100~keV for imaging \cite{GOTZFRIED2018286}.
Finally, because photonuclear reaction cross sections typically exhibit a giant dipole resonance between 10~MeV and 30~MeV, photoneutron emission is favored in this energy regime, which enables the efficient generation of intense, short-pulse sources for neutron spectroscopy \cite{yogo2023advances}
For all these applications, high peak flux and particle yield are critical, particularly when probing HED regimes driven by low-repetition-rate, high-power lasers.
Therefore, the development of LWFA platforms capable of generating high-peak-flux particle and X-ray backlighters using ultra-intense (sub)picosecond lasers has drawn increasing interest over the past decade, particularly for integration into large-scale HED facilities like the National Ignition Facility (NIF), OMEGA, XG-III, GEKKO XII or the Laser Mégajoule (LMJ). However, realizing these capabilities requires overcoming technical challenges, such as designing specialized targets and diagnostics that can operate in harsh, high-power HED environments while complying with strict facility constraints.

In LWFA, an intense laser pulse propagates in an underdense plasma and drives a trailing plasma wave, which takes the form of co-propagating ion cavities \cite{tajima1979laser,esarey2002overview}.
Electrons can be trapped into one or several of these cavities where they experience strong longitudinal electric fields (10's to 100's MeV/cm), thus reaching relativistic energies over very short (mm to cm) distances. 
In the blowout regime, associated with femtosecond laser pulses shorter than the plasma period, quasi-monoenergetic electron beams can be produced with an energy up to the 10 GeV scale \cite{Gonsalves_PRL_2019, Aniculaesei_MRE_2023, Picksley_PRL_2024} and a typical charge ranging from a hundred pC up to a few nC \cite{pukhov2002laser, PhysRevX.10.041015, Aniculaesei_MRE_2023}. By contrast, when the laser pulse duration is longer than the plasma period, the laser envelope self-modulates at the plasma period under the combination of longitudinal and/or transverse instabilities such as Raman forward scattering, relativistic self-phase modulation and envelope self-modulation, which leads to effective wakefield excitation \cite{Joshi_PRL_1981,andreev1992resonant, Antonsen_PRL_1992, Sprangle_PRL_1992, esarey1994envelope, Mori_IEEE_JQE_2002}. This regime is called self-modulated LWFA (SMLWFA) \cite{PhysRevLett.78.3125, PhysRevLett.86.1227, modena1995electron, sprangle1988laser, Ferri_PRAB_2016, Albert_PRL_2017, Li_HEDP_2020, Sha21}.
While the resulting maximum electron energy in this regime -- reaching up to 380 MeV as reported in \cite{Lemos_PPCF_2018, Lemos2019} -- remains more modest compared to the blowout regime, and the electron beam quality is lower (exhibiting a Maxwellian-like spectrum and higher divergence), significantly higher charges can be achieved. Indeed, charges of up to $\sim$700~nC have been demonstrated via SMLWFA \cite{Sha21}, which can be explained by (i) the use of picosecond-long laser pulses with energies of the order of hundreds of Joules, (ii) the high laser-to-electron conversion efficiencies (typically several percent in this regime, favored by electron trapping over multiple plasma periods), and (iii) the moderate average electron energies (tens of MeV). 

Blowout LWFA and SMLWFA typically occur at moderate laser intensities ($\sim 10^{19}\,\rm W\,cm^{-2}$) and low plasma densities (below 1\% of the critical density, $n_c=m_e\epsilon_0\omega_0^2/e^2$ where $\omega_0$ is the laser radial frequency, $m_e$ and $e$ are the electron mass and charge, respectively, and $\epsilon_0$ is the permittivity of free space). 
At higher -- yet still undercritical --  plasma densities ($\sim 0.1\,n_c$), the laser drives a plasma channel that generates focusing forces rather than a wakefield. The coupling between the transverse electron oscillations in the plasma channel and the laser field can then lead to direct laser acceleration (DLA) \cite{pukhov1999particle, Khudik_POP_2016, Hussein_NJP_2021, Babjak_PRL_2024, Babjak_NJP_2024, Tang_NJP_2024, Tang_POP_2025, Gyrdymov_SR_2024}, which also typically yields Maxwellian spectra and high beam divergence. Although electron energies are generally lower in this regime due to the higher plasma densities, a record energy of 500 MeV has been reported\cite{Hussein_NJP_2021}.
At intermediate densities (close to $0.01~n_c$), both wakefield acceleration and DLA -- induced by the focusing field of the wakefield and the resulting betatron motion -- can occur. Experimental evidence, supported by simulations, of DLA occurring on top of LWFA has been obtained for short ($<50\,\rm fs$) laser pulses\cite{Shaw_PRL_2017, Gallardo_Gonzalez_NJP_2018, Shaw_PPCF_2018}.
In the SMLWFA regime, the inherent overlap of the long laser pulse with electrons accelerated and oscillating in the wakefield results in a complex interplay between wakefield acceleration and DLA\cite{mangles2005electron, Albert_PRL_2017, Li_HEDP_2020, Sha21, King_PRL_2021, Lemos_PRR_2024}. Since the electron beam properties of SMLWFA and DLA are difficult to distinguish experimentally, only detailed particle-in-cell (PIC) simulations, capable of separating laser and plasma fields, can quantify their respective contributions \cite{King_PRL_2021, Miller_POP_2023}.
For example, at a density near $0.04\,n_c$ -- a regime typically favorable for DLA -- the use of a high-intensity ($>10^{21}\,\rm W\,cm^{-2}$), short-pulse ($\sim 25\,\rm \rm fs$) PW-class laser was shown to produce a Maxwellian spectrum with maximal energies up to tens of MeV, high charge ($\sim 250\,\rm nC$), yet with LWFA as the dominant acceleration mechanism in this case, according to PIC simulations \cite{Hu_APR_2025}.

In this paper, we present the results of experiments carried out at LMJ making use of the kilojoule-class sub-picosecond PETAL laser beam in a mixed regime of SMLWFA and DLA. An electron beam with a Maxwellian-like distribution was generated, exhibiting a maximum energy of approximately 500 MeV, with a total charge exceeding $1\,\mu \rm C$ for energies above 1~MeV. When interacting with a thick, high-$Z$ target, this beam produced a multi-MeV, sub-picosecond, Joule-class X-ray flash via Bremsstrahlung, achieving a laser-to-photon conversion efficiency of up to 0.5\%. Comprehensive numerical simulations were performed to better estimate uncertainties in the reported values and calibrate the gas jet and detector parameters. Furthermore, 3D and quasi-3D PIC simulations allowed us to unravel the respective contributions of SMLWFA and DLA mechanisms and their interplay. These results demonstrate that our SMLWFA/DLA platform, which integrates a supersonic gas jet illuminated by the PW-class PETAL laser beam within the same experimental chamber where the high-energy nanosecond LMJ beams are focused, offers a unique capability for future pump-probe experiments.

\section*{Results and discussion}

\subsection*{Experimental setup}

\begin{figure}[htbp]
\centering
\includegraphics[width=1\textwidth,trim={1.5cm 5cm 0cm 0},clip]{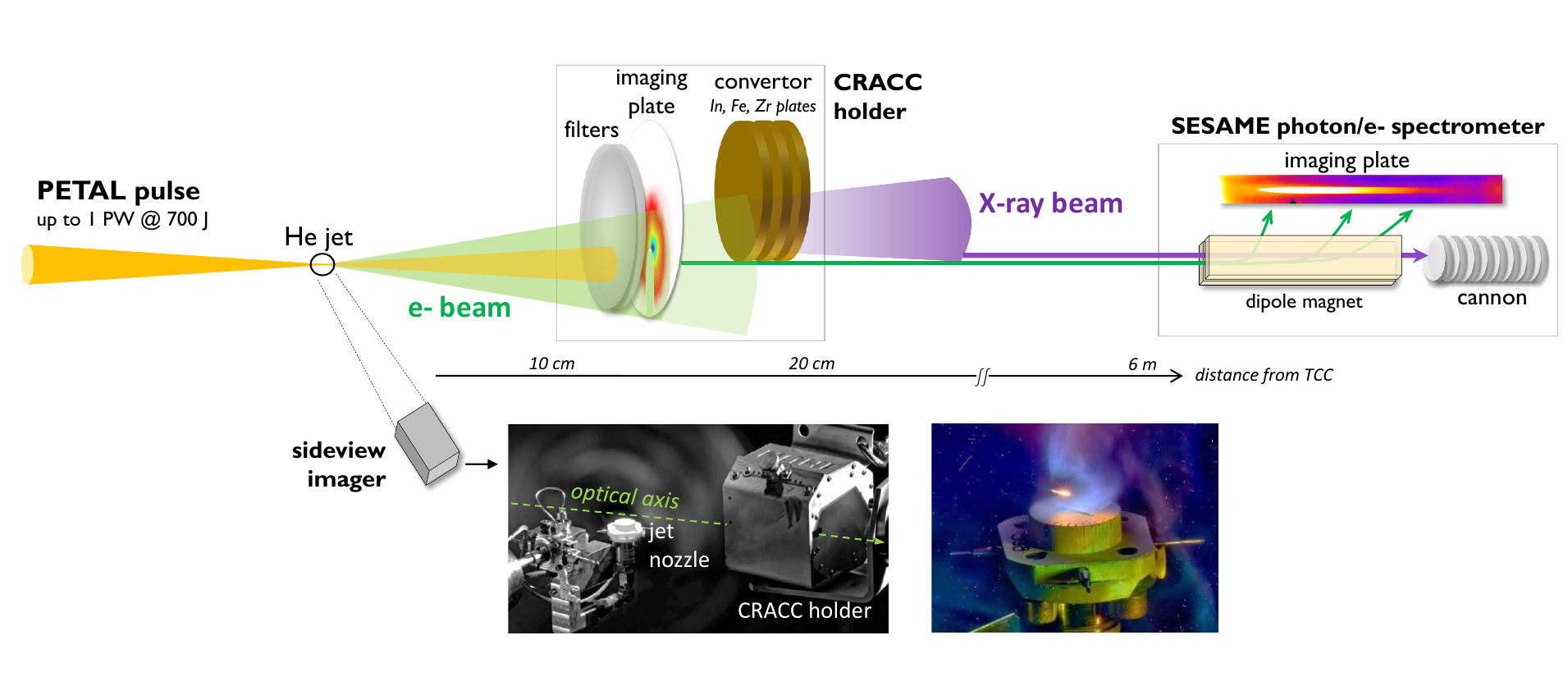} 
\caption{(top) Top-view schematic of the experimental setup and (bottom) side-view images of the target chamber center (TCC) region recorded before (left) and during (right) the shot (see text for details).}
\label{fig:setup}
\end{figure}

The experiment was performed using the PETAL laser system at the LMJ facility \cite{Cay24,Neauport:24}. The setup is depicted in Fig.~\ref{fig:setup}. PETAL is a PW-class, Nd:glass chirped-pulse amplified laser chain \cite{strickland1985compression} with a central wavelength of 1053 nm, a minimum pulse duration of $\sim 700\,\rm fs$ and a maximum pulse energy currently of $\sim 700\,\rm J$. Laser pulse duration, focal spot and energy were monitored at each shot. The PETAL beam was focused in a $f/20$ focusing geometry at the target chamber center (TCC) of the LMJ experimental chamber, where it interacted with a supersonic He gas jet. The jet was delivered by a Parker gated valve coupled with a gas nozzle produced by additive layer manufacturing. Two different gas nozzles were used: a standard Laval-type circular nozzle with a 4\,mm output diameter and a 10\,mm$\times$1\,mm slit aperture nozzle. A maximum gas pressure of 60~bars could be reached by the regulation system. The effective gas density in the interaction area was estimated by means of computational fluid dynamics simulations \cite{openfoam} and offline Mach-Zehnder interferometric measurements, resulting in an electron plasma density of up to 2--$3 \times 10^{19}\,\rm^{-3}$ in the resulting plasma (see Methods).

The accelerated electrons were collected by a MS-type imaging plate (IP) placed in the so-called CRACC (Cassette for RAdiography at Chamber Center) holder. A set of three 1~cm-thick plates of natural indium, iron and zirconium with 75-mm diameter was inserted behind the IP. These plates acted as a Bremsstrahlung convertor producing a beam of hard X-rays in the MeV range. 
To allow the on-axis beam electrons to propagate towards the spectrometer, an aperture was made in the IP and in the frame of CRACC, and the convertor was placed slightly off-axis.
Part of the electron beam could then reach the SESAME (Spectrometer for ElectronS At Moderate Energies) device, located 6.18 m away from the TCC, at the edge of the experimental chamber.
In SESAME, the electron beam spectrum was measured by means of a static dipole magnet deflecting the electrons onto one or several IP(s). A ``cannon stack'' spectrometer, positioned behind the dipole, charaterized the hard X-ray spectrum. The stack was composed of alternating layers of IPs and metallic filters with increasing atomic number and/or thicknesses to provide energy resolution \cite{Chen2008ABS,Koester2020BremsstrahlungCD}.
Finally, a camera coupled to a long working distance objective was used for sideview imaging. Thanks to a sufficiently long aperture time, it could capture the scattered light signal in the visible range over the entire duration of the gas flow ($\sim$10~ms), thus including the laser-plasma interaction time.

\begin{table}[htbp]
\centering
\begin{tabular}{|c|c|c|c|c|}
\hline
\multirow{2}{*}{shot}  & \multirow{2}{*}{nozzle}  & peak intensity & pulse duration & energy
\\
 & & [$\rm W\,cm^{-2}$] & [ps] & [J] 
\\
\hline
1  & conical 4 mm  & 4.4$\times$10$^{18}$ & 0.64 & 347 \\
2  & conical 4 mm  & 4.3$\times$10$^{18}$ & 0.915 & 313  \\
3  & slit 10 mm  & 7.9$\times$10$^{17}$ & $\sim$ 4 & 352   \\
4  & slit 10 mm  & 4.6$\times$10$^{18}$ & 0.82 & 351  \\
5  & slit 10 mm  & 2.0$\times$10$^{19}$ & 0.753 & 719  \\
\hline
\end{tabular}
\caption{Shot list with associated target and laser parameters.}
\label{tab:shots}
\end{table}

\begin{figure}[htbp]
\centering
\begin{subfigure}{0.49\textwidth}
    \includegraphics[width=\textwidth]{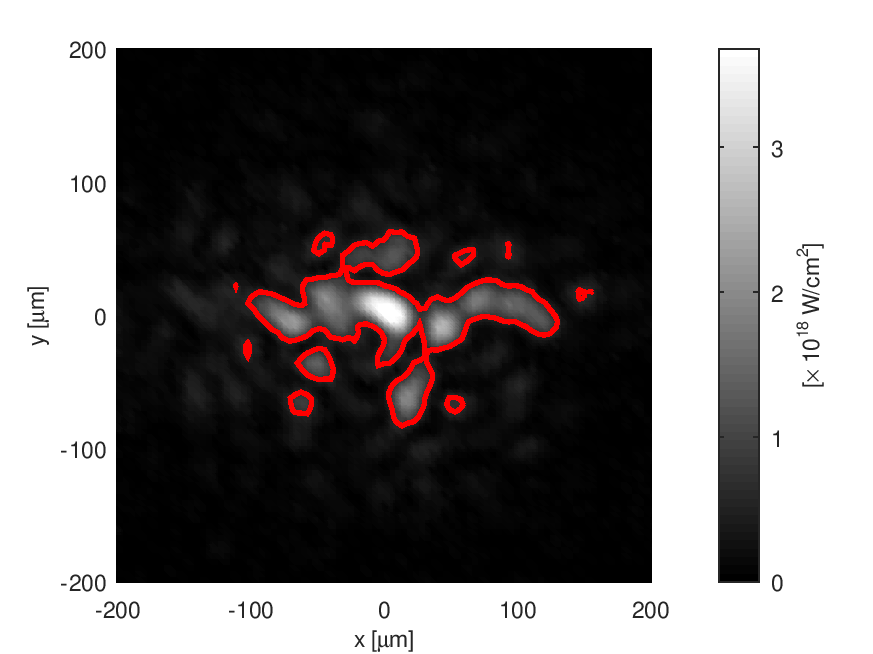}
    \caption{}
    \label{fig:tachea}
\end{subfigure}
\begin{subfigure}{0.49\textwidth}
    \includegraphics[width=\textwidth]{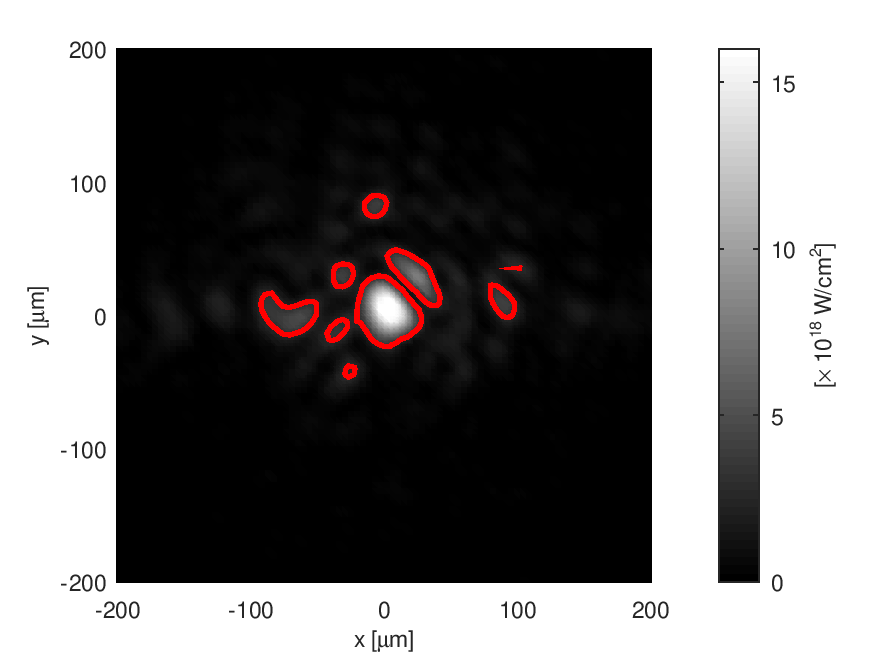}
    \caption{}
    \label{fig:tacheb}
\end{subfigure}
\caption{Measured PETAL focal spot for shot~\#4 (a) and shot~\#5 (b). The encircled areas correspond to portions above the $1/e^2$  threshold. The focal spot is measured at each shot across the entire beam aperture, on a diagnostic line located at the compressor output.}
\label{fig:tache}
\end{figure}

In the following, we present the results of five laser shots, the main characteristics of which are reported in Table~\ref{tab:shots}. The backing pressure of the jet was similar for all the shots (50--60~bars). In the first four shots, the laser energy was limited to $\sim$350~J. For the last shot the laser chain was upgraded to provide an improved focal spot quality, a total energy of 719~J and a peak power of 0.95~PW. Representative measured PETAL focal spots are shown in Fig.~\ref{fig:tache}. The gray level indicates the intensity value that is determined locally (i.e., at every pixel of the detector), assuming that the pulse duration is homogeneous over the transverse distribution. The red contours encompass the portions of the focal spot above a threshold taken at $1/e^2$ of the peak. The contours show a significant improvement of the focal spot quality for shot~\#5 (Fig.~\ref{fig:tacheb}) in comparison to the other shots (e.g., Fig.~\ref{fig:tachea} corresponding to shot~\#4).
For this shot, the peak intensity on target reached $2.0\times10^{19}\,\rm W.cm^{-2}$. It reduces to an average of $\sim 8$--$9\times 10 ^{18}\,\rm W\,cm^{-2}$ within the central hotspot and to an average of $5.8 \times 10^{18}\,\rm W\,cm^{-2}$ in the regions above the $\frac{1}{e^2}$ threshold, containing $27\%$ of the total shot energy. 

\subsection*{Electron beam}

\begin{figure}[htbp]
\includegraphics[width=\textwidth]{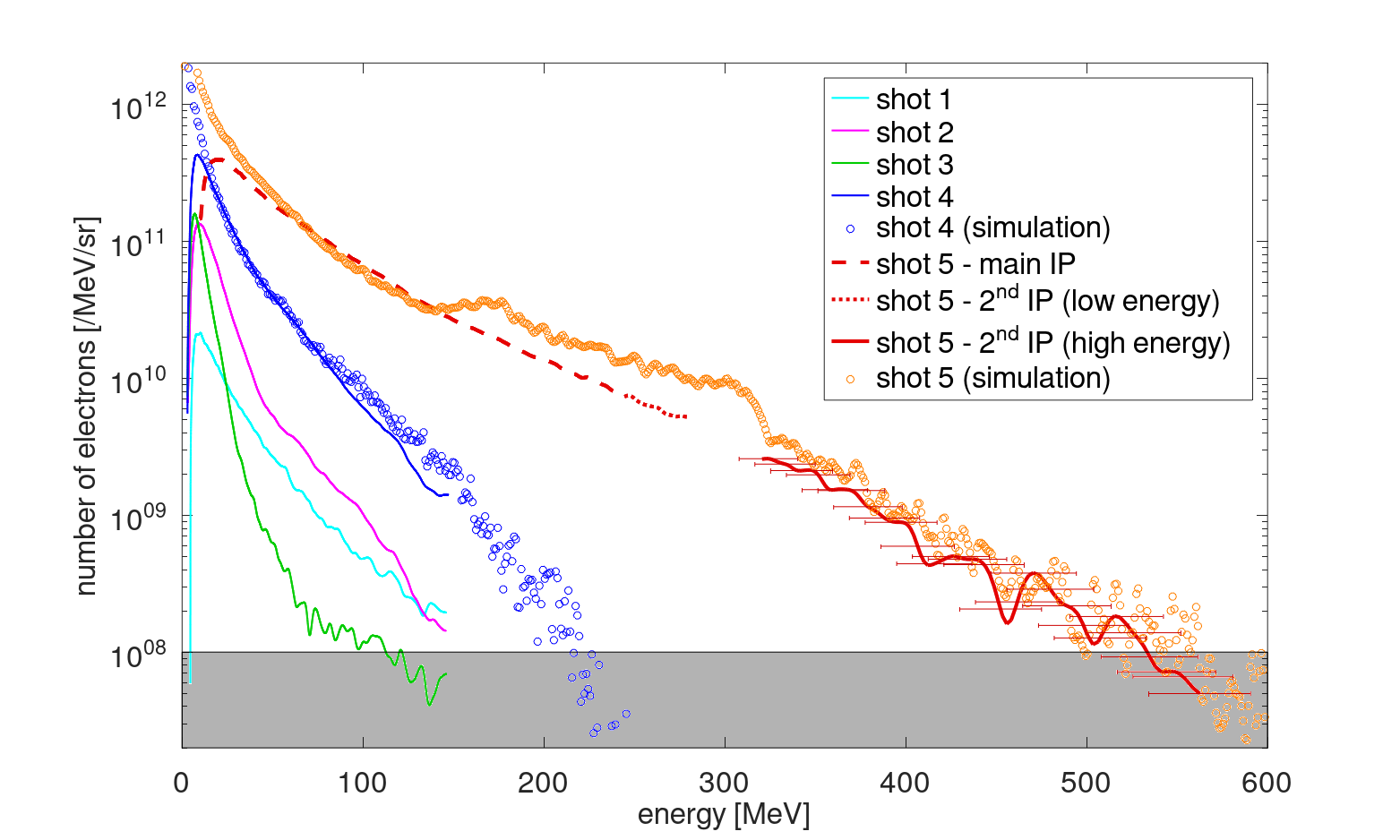}
\caption{Measured (lines) and simulated (circles) on-axis electron beam spectra. For shots~\#1--4, the high-energy detection limit of SESAME was $\sim 150\,\rm MeV$, whereas for shot~\#5 it was extended beyond $500\,\rm MeV$. Horizontal error bars corresponding to the energy resolution indicate the uncertainty in the maximum detected energy for shot~\#5. The gray shaded area represents the average detection threshold of SESAME ($\sim 10^8\,\mathrm{e^-/MeV/sr}$).}
\label{fig:spectra}
\end{figure}

The electron beam spectra for all five shots, measured on-axis by SESAME, are shown in Fig.~\ref{fig:spectra}. The fall at low energies ($E \lesssim 15~\rm MeV$) is not representative of the effective spectrum, but is due to the detection limit of the spectrometer. Above this threshold, the spectrum is well fitted by a two-temperature Maxwellian distribution, from which we calculated the weighted mean energy of the electron beam.

For the first four shots, a maximum mean energy of 23~MeV was obtained for the shortest-pulse, short-nozzle case (shot~\#1). A comparable energy distribution, though at a much higher magnitude, was recorded for shot~\#4, which corresponds to a similar laser intensity but used the long (10-mm slit) nozzle. Note that the higher spectrum amplitude observed in shot~\#4 may also be attributed to laser fluctuations in addition to the nozzle change. 
For this first series, the maximum detectable energy was approximately of 150~MeV, which prevented the high-energy tail of the electron spectrum from being measured, especially for shot~\#4. We thus modified the SESAME configuration for the last shot in order to extend the detection limit (see details in Methods).
On this last shot, maximizing the PETAL intensity allowed us to boost the mean electron energy to 50~MeV, and maximum energies up to 500--550~MeV could be observed. Here, the resolution in the high-energy region of the spectrum was limited to $\Delta E / E\sim 0.05$, mainly determined by the 5-mm entrance slit of SESAME. The curves marked with circles in Fig.~\ref{fig:spectra} show the results of three-dimensional PIC simulations performed with the CALDER code~\cite{Lefebvre_NF_2003}. These simulations were carried out using laser and plasma parameters corresponding to shots~\#4 and \#5 (see simulation details in Methods).

\begin{figure}[htbp]
\centering
\begin{subfigure}{0.19\textwidth}
    \includegraphics[width=\textwidth]{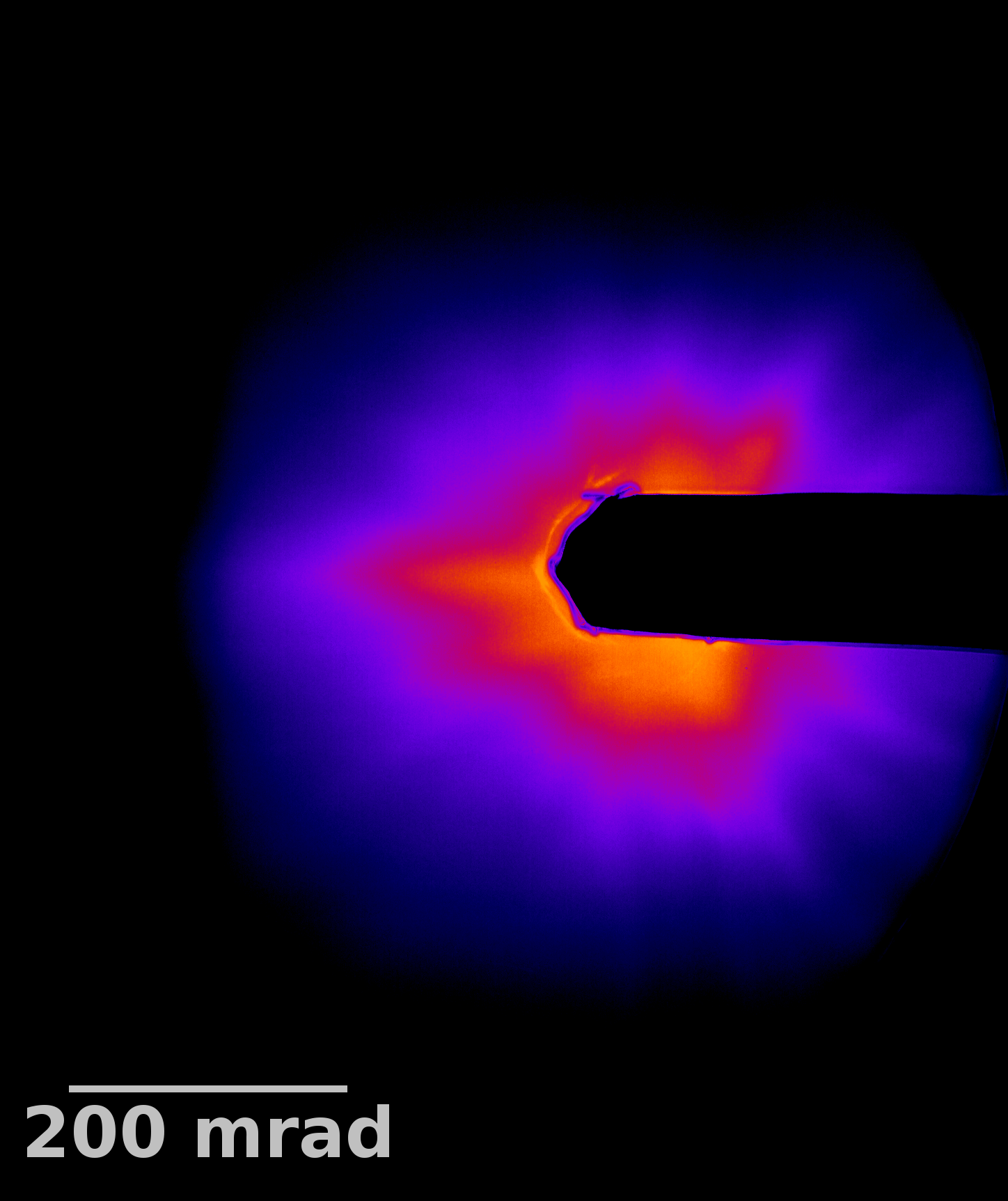}
    \caption{shot 1}
    \label{fig:ebeama}
\end{subfigure}
\begin{subfigure}{0.19\textwidth}
    \includegraphics[width=\textwidth]{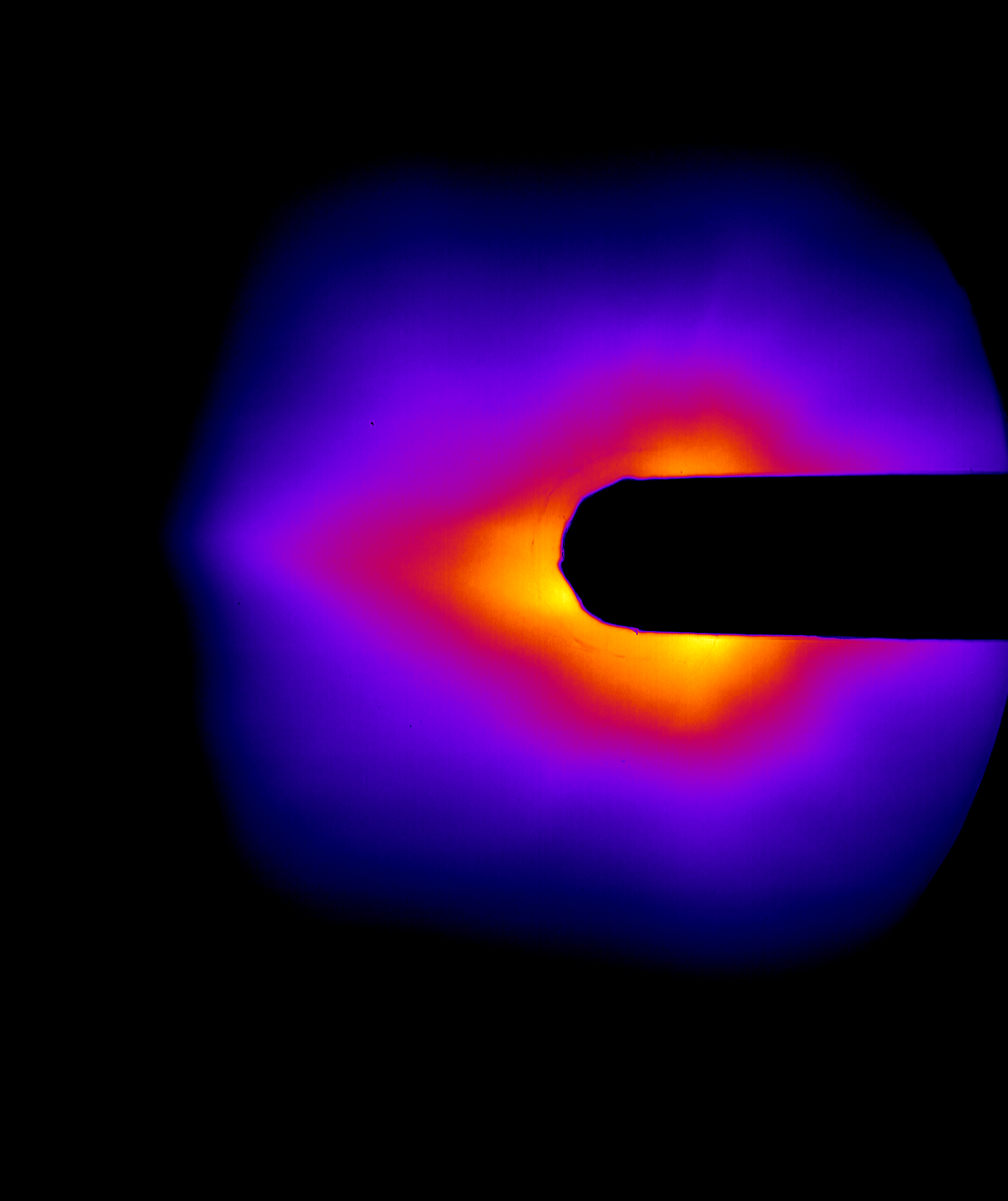}
    \caption{shot 2}
    \label{fig:ebeamb}
\end{subfigure}
\begin{subfigure}{0.19\textwidth}
    \includegraphics[width=\textwidth]{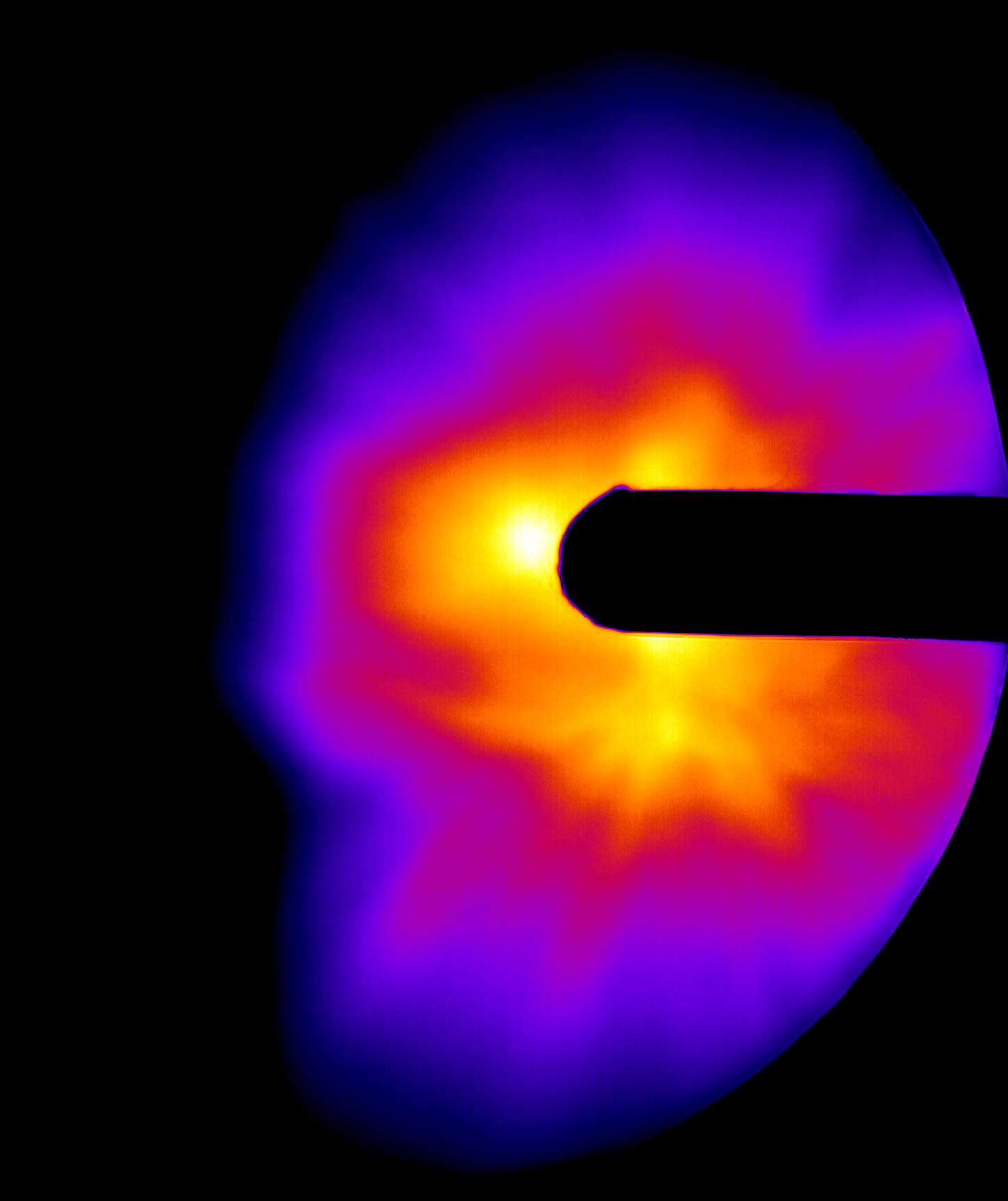}
    \caption{shot 3}
    \label{fig:ebeamc}
\end{subfigure}
\begin{subfigure}{0.19\textwidth}
    \includegraphics[width=\textwidth]{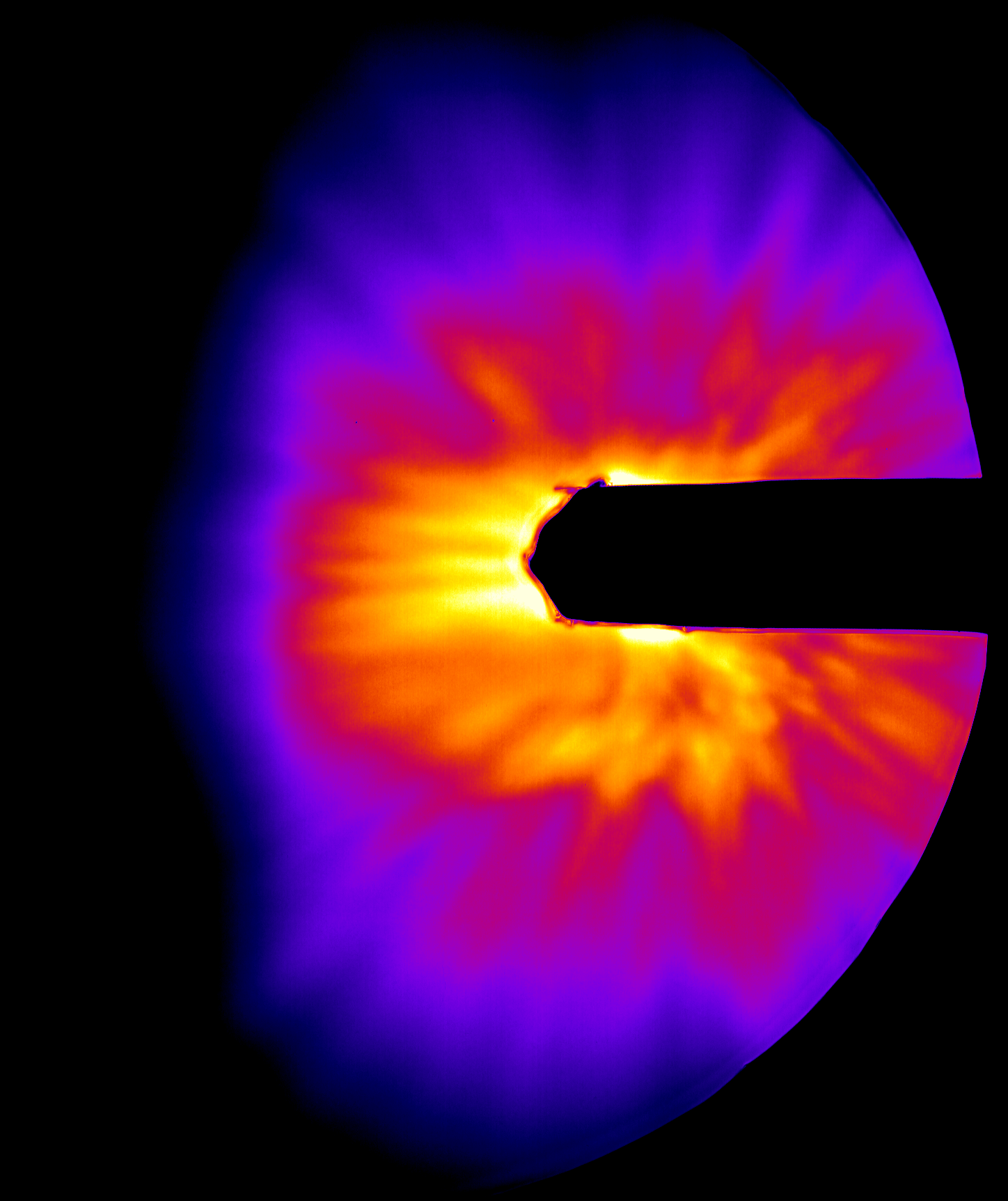}
    \caption{shot 4}
    \label{fig:ebeamd}
\end{subfigure}
\begin{subfigure}{0.19\textwidth}
    \includegraphics[width=\textwidth]{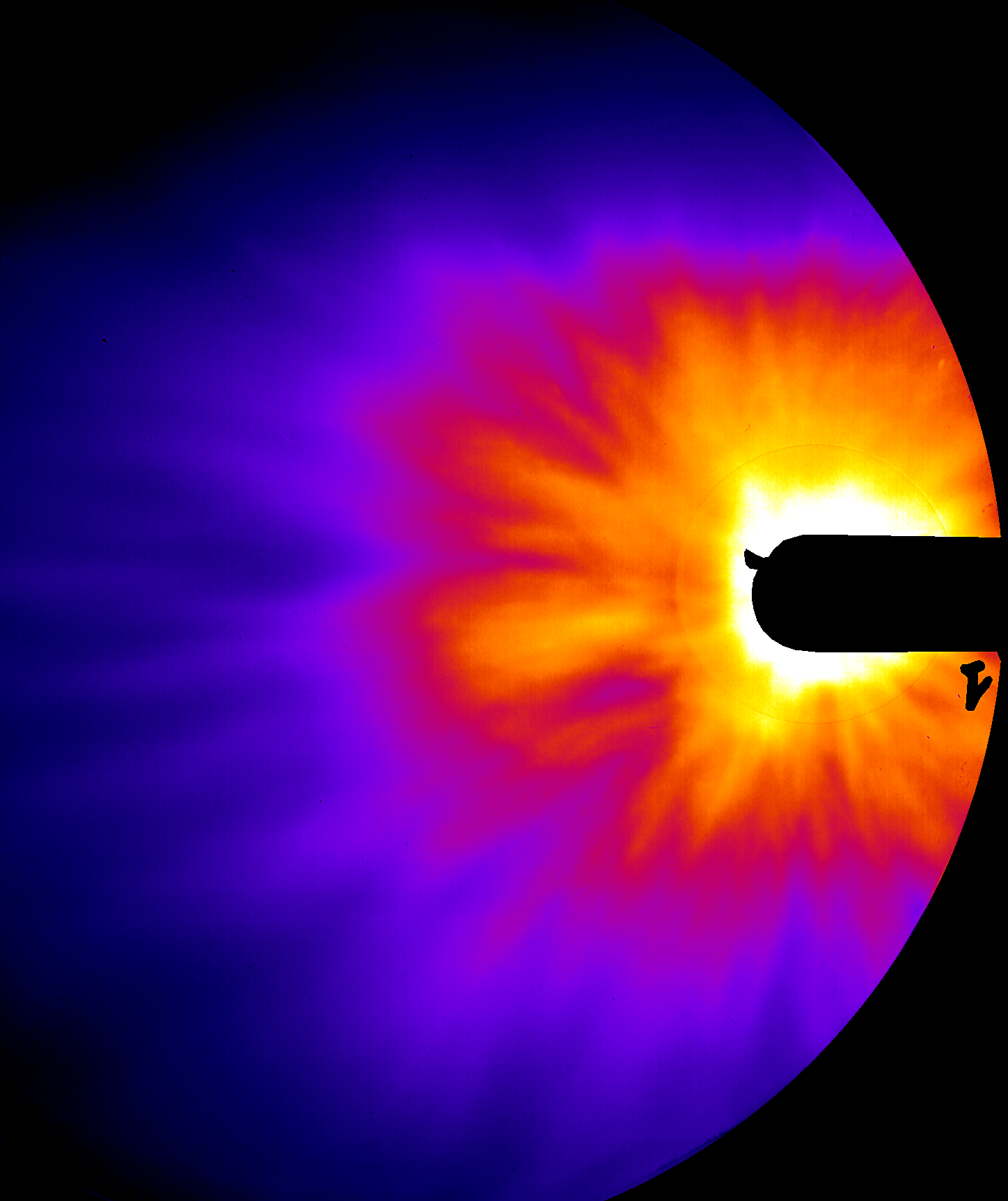}
    \caption{shot 5}
    \label{fig:ebeame}
\end{subfigure}
\caption{(a--e) Electron beam profiles measured on the CRACC IP for the five reported shots. For visual clarity, the color scale, which represents the raw photostimulated luminescence (PSL) signal recorded after scanning (see Methods), is arbitrary and normalised independently for each shot. The spatial scale bar is shown in the lower-left corner of panel (a). The missing signal in the central region corresponds to the aperture drilled through the IP and CRACC frame to allow electrons to propagate toward SESAME.}
\label{fig:ebeam}
\end{figure}

Figure~\ref{fig:ebeam} shows the spatial electron beam profiles as measured by the CRACC IP detector.
The total charge impacting the IP is inferred from its known calibration~\cite{Bou15}. Filtering layers placed in front of CRACC set a low-energy threshold for electrons reaching the detectors; specifically, only electrons with $E \gtrsim 0.9\,\rm MeV$ could reach the first IP. Consequently, all charge values reported here correspond strictly to electrons above this energy threshold.

For better accuracy, we carried out Monte Carlo Geant4 simulations incorporating the full CRACC geometry to determine the response functions of the IPs within our experimental configuration. This allowed us to model the secondary background signal generated within CRACC and isolate the primary electron beam contribution in the charge analysis. Because the IP response varies with electron energy (despite remaining nearly constant above a few MeV) and the spectrum varies shot-to-shot, customized response functions were computed for each shot by spectrally averaging the responses using the electron spectra reported in Fig.~\ref{fig:spectra} as input.

The highest charge was obtained using the 10--mm slit nozzle at maximal PETAL intensity (shot~\#5). For this shot, corresponding to the beam profile shown in Fig.~\ref{fig:ebeame}, we achieved a charge value of $1.1 \pm 0.13\,\mu\rm C$. To estimate the signal loss caused by the IP’s central hole, we performed a two-dimensional fit to the surrounding signal, which we extrapolated into the central region. Even when neglecting the contribution from this central area, the raw charge collected for this shot reaches $0.97\,\mu\rm C$. The estimated beam divergence--defined as the half-width at half-maximum (HWHM) of Lorentzian fits to the horizontal and vertical electron beam profiles--is $209 \times 152\,\rm mrad^2$. A fraction of $31\%$ of the total charge ($\sim 339\,\rm nC$) is contained within the FWHM area. Table~\ref{tab:chargeresults} summarizes the charge measurements across all reported shots, and the complete IP analysis procedure is detailed in the Methods section.

\begin{table}[htbp]
\centering
\begin{tabular}{|c|c| >{\bfseries}c|}
\hline
\multirow{2}{*}{shot} & raw collected charge & total beam charge
\\
 &  [nC] & [nC] \\
\hline
1 & 58 $\pm$ 43 & 72 $\pm$ 56 \\
2 & 170 $\pm$ 38 & 219 $\pm$ 52  \\
3 & 548 $\pm$ 67 & 658 $\pm$ 92 \\
4 & 457 $\pm$ 105 & 597 $\pm$ 145  \\
5 & 974 $\pm$ 114 & 1094 $\pm$ 128  
\\
\hline
\end{tabular}
\caption{Summary of charge measurements across all shots. Second column: raw charge measured onto the active area of the IP, obtained following the procedure described in the Methods section. Third column: total charge value obtained after extrapolation of the missing signal in the central area, integrated over an area bounded by the second-moment beam diameters. All values correspond to electron energies higher than 0.9~MeV.}
\label{tab:chargeresults}
\end{table}

\begin{figure}[htbp]
\centering
\begin{subfigure}{0.49\textwidth}
\includegraphics[width=\textwidth]{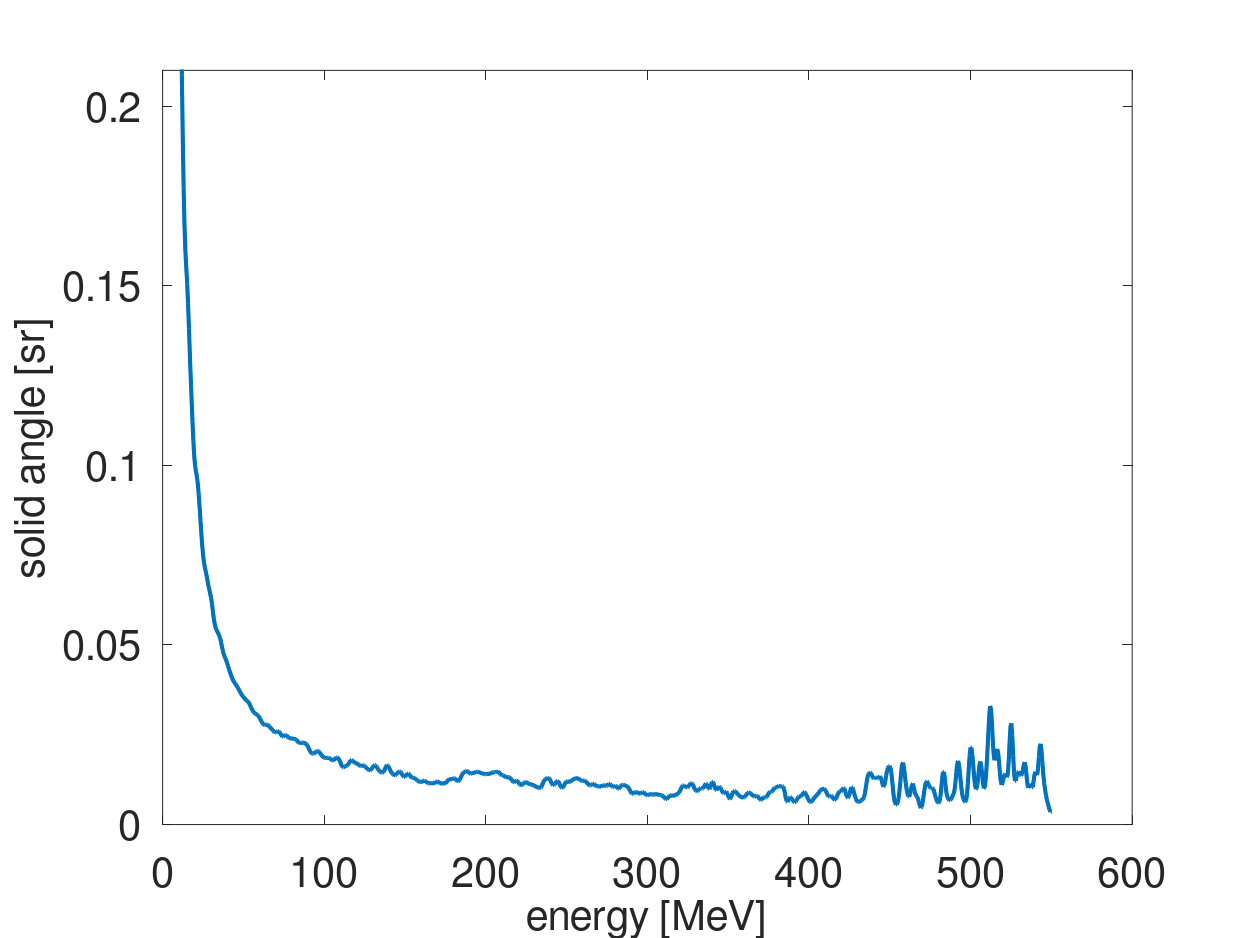}
\caption{}
\label{fig:spsimusa}
\end{subfigure}
\begin{subfigure}{0.5\textwidth}
\includegraphics[width=\textwidth]{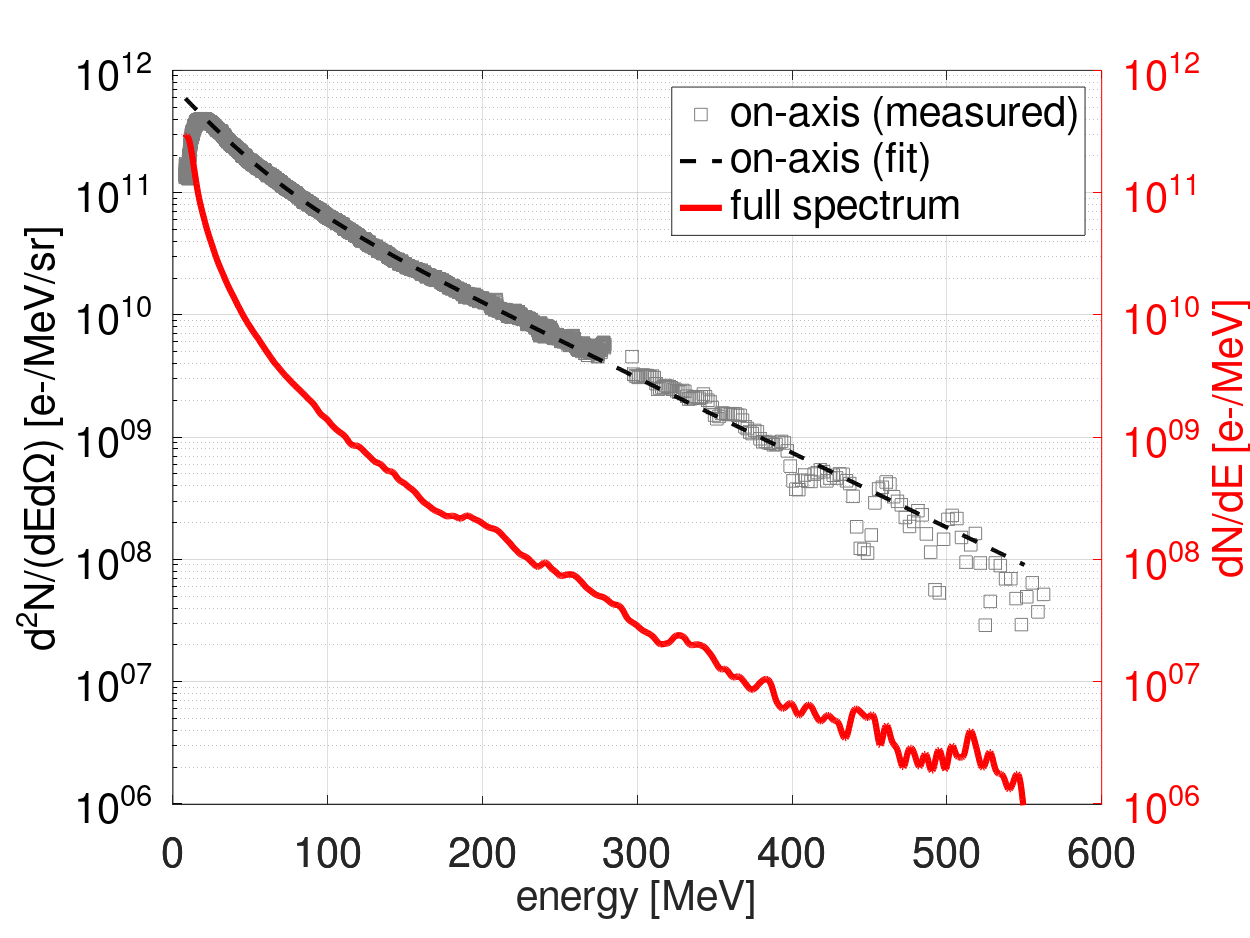} 
\caption{}
\label{fig:spsimusb}
\end{subfigure}
\caption{a) RMS solid angle vs. electron beam energy, $\Omega_{\rm sim}(E)$, as extracted from a 3D PIC simulation of shot~\#5. (b) Corresponding energy spectrum $dN/dE$ integrated over the full transverse beam distribution (solid red line), compared to the on-axis spectrum $d^2N/(dE d\Omega)$ observed by SESAME (squares: raw measurement; dashed line: two-temperature fit $S(E)$). The total spectrum is calculated as $dN/dE = S(E) \cdot \Omega_{\rm sim}(E)$. Note that integrating the $dN/dE$ curve yields a total charge of $1.06\,\rm \mu C$, in excellent agreement with the charge directly measured by CRACC ($1.1\,\rm \mu C$).}
\label{fig:spsimus}
\end{figure}

The conversion efficiency from laser energy to total energy carried by the electron beam is given in practical units by $\frac{\text{charge [}\mu\text{C}] \times \text{mean energy [MeV]}}{\text{laser energy [J]}}$. However, the mean energy extracted from the SESAME measurement corresponds strictly to the on-axis beam distribution (captured within an acceptance angle $<1\,\rm mrad$), which is not representative of the true transverse electron distribution spanning 100s of mrad. Although the latter could not be measured experimentally, it can be inferred from PIC simulations, as shown in Fig.~\ref{fig:spsimusa} for shot~\#5. Below $\sim 50\,\rm MeV$, the root-mean-square (RMS) solid angle of the electrons increases significantly. Consequently, the most energetic electrons contribute very little to the off-axis signal, meaning that the mean energy weighted over the full 2D transverse distribution is lower than that measured on-axis. Because the PIC simulations accurately reproduce the experimental on-axis electron spectrum (Fig.~\ref{fig:spectra}), we can reliably use the PIC results to deduce the average energy of the entire beam. The total energy spectrum, $dN/dE$, was thus reconstructed for shot~\#5 using the simulated energy–angle distribution (Fig.~\ref{fig:spsimusa}) together with the experimental on-axis spectrum (Fig.~\ref{fig:spectra}). The reconstructed spectrum in Fig.~\ref{fig:spsimusb} yields a full-beam mean energy of $15\,\rm MeV$, compared to $50\,\rm MeV$ measured on-axis. Taking into account the total laser energy of $719\,\rm J$, the corresponding laser-to-electron conversion efficiency is $2.3\%$.

However, because the wings of the PETAL focal spot may not participate in relativistic electron acceleration, we can restrict our consideration to the area above the $1/e^2$ threshold (Fig.~\ref{fig:tacheb}), which provides the useful energy effectively involved in the acceleration process. The corresponding encircled laser energy is $E_{\rm use} = 194\,\rm J$, yielding an effective conversion efficiency of $8.5\%$. 
This is comparable to the highest values reported for SMLWFA~\cite{Sha21}, which were notably obtained by considering only the more energetic, on-axis spectrum. Improving the spatial quality of the PETAL focal spot on target thus represents a promising route toward achieving even higher conversion efficiencies.

\subsection*{Self-modulation dynamics and contribution of DLA}

As shown in Fig.~\ref{fig:spectra}, 3D CALDER PIC simulations reproduce the measured on-axis spectrum well for shots~\#4 and \#5, helping infer the total angle-integrated spectra from the measured on-axis data (Fig.~\ref{fig:spsimus}).

The simulation of shot~\#4 was investigated in greater detail. In this run, the initial laser intensity was $4.6 \times 10^{18}\,\rm W\,cm^{-2}$, yielding a peak vector potential $a_0 = 1.825$ (normalised to $m_e c/e$, where $m_e$, $e$, and $c$ are the electron mass, elementary charge, and speed of light, respectively). The experimental focal spot shown in Fig.~\ref{fig:tachea} was approximated as the coherent sum of two Gaussian beams to capture both the central spot elongation and the energy in the wings. The first beam, modeling most of the central hotspot, is defined by $a_1 = 1.267$ with focal spot sizes $w_{y1} = 24.2\,\rm \mu m$ (full-width at half-maximum, FWHM, of the transverse intensity profile) and $w_{x1} = 39.9\,\rm \mu m$. The second beam, capturing the energy in the wings, is characterized by $a_2 = 0.558$, $w_{y2} = 388\,\rm  \mu m$, and $w_{x2} = 484.1\,\rm \mu m$, yielding a total amplitude $a_0 = a_1 + a_2$. The laser pulse duration is $820\,\rm fs$ (FWHM of the temporal intensity profile) with a total energy of $351\,\rm J$. The transverse size of the simulation box is $214\,\rm \mu m$, cutting off part of the wings so that $\sim 200\,\rm J$ only enters the box. The plasma density profile reproduces the 10--mm nozzle gas jet shown in Fig.~\ref{fig:gasdensity}.

\begin{figure}[htbp]
\centering
\begin{subfigure}{0.56\textwidth}
    \includegraphics[width=\textwidth]{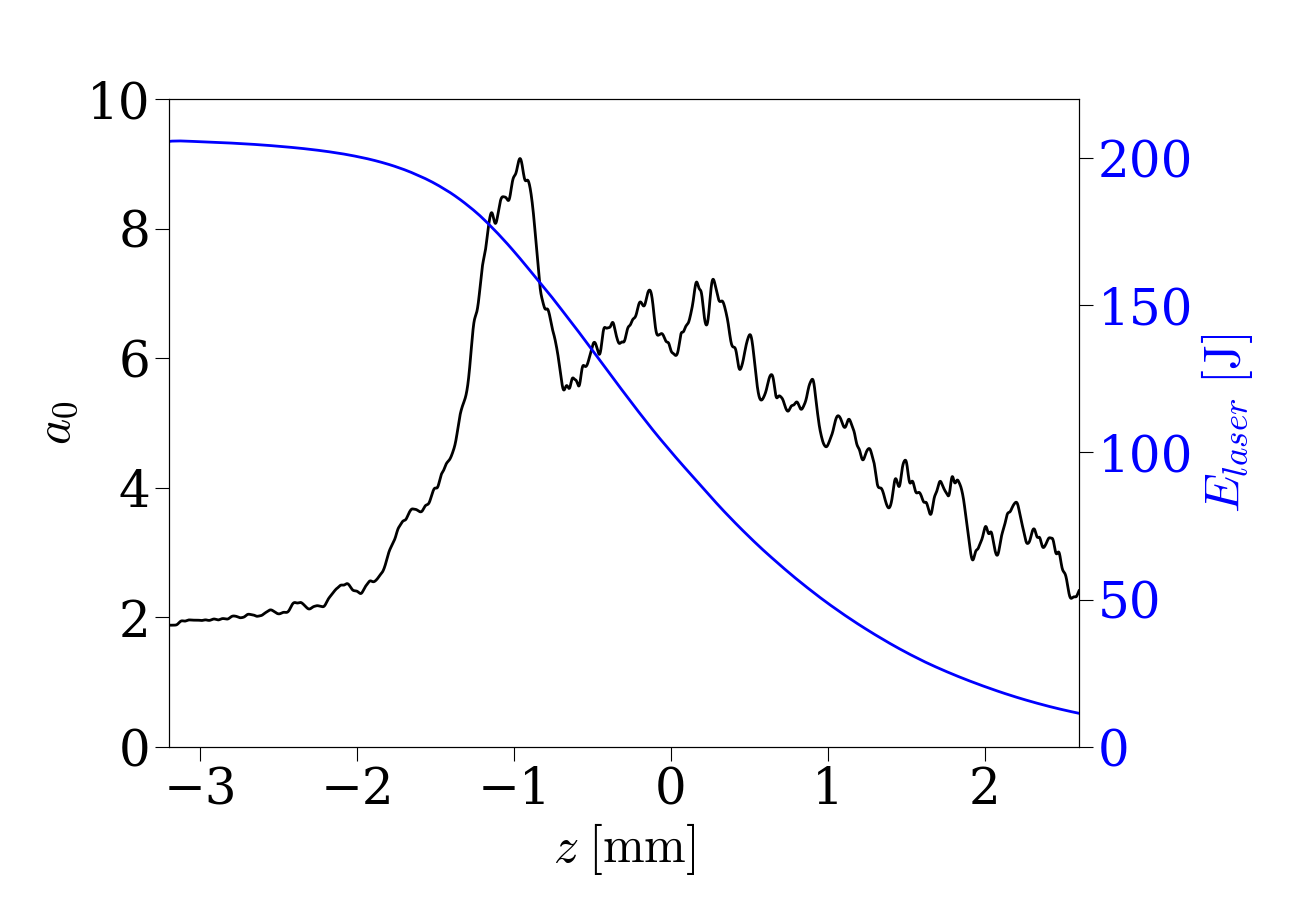}
    \caption{}
    \label{fig:sim3Da0}
\end{subfigure}
\begin{subfigure}{0.33\textwidth}
    \includegraphics[width=\textwidth]{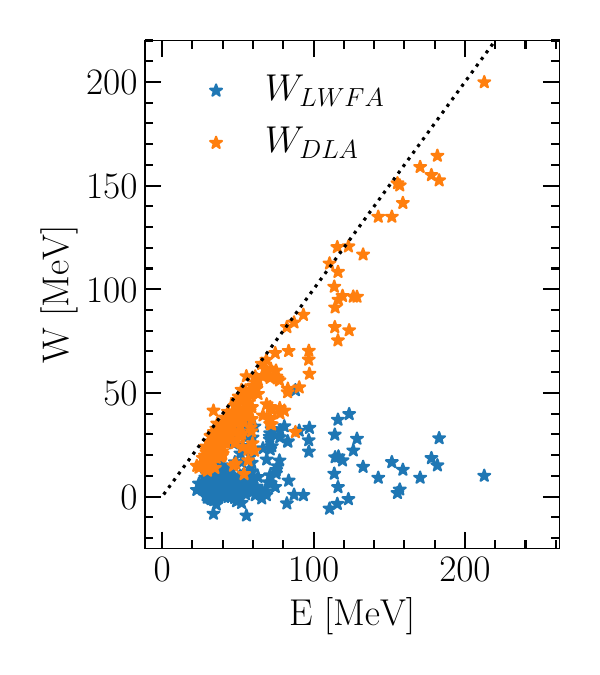}
    \caption{}
    \label{fig:simOsiris}
\end{subfigure}
\begin{subfigure}{0.33\textwidth}
    \includegraphics[width=\textwidth]{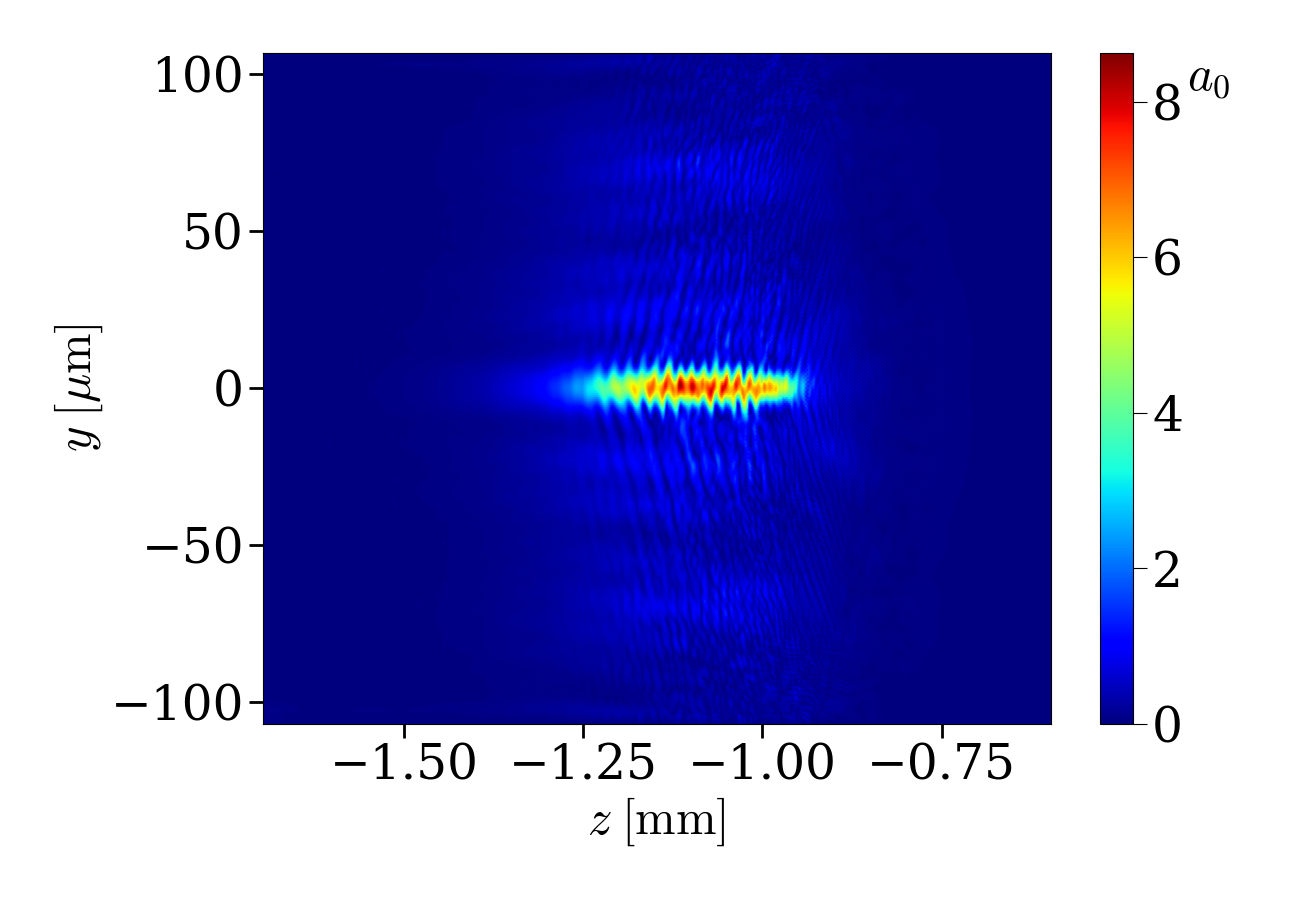}
    \caption{}
    \label{fig:sim3Dlas1}
\end{subfigure}
\begin{subfigure}{0.33\textwidth}
    \includegraphics[width=\textwidth]{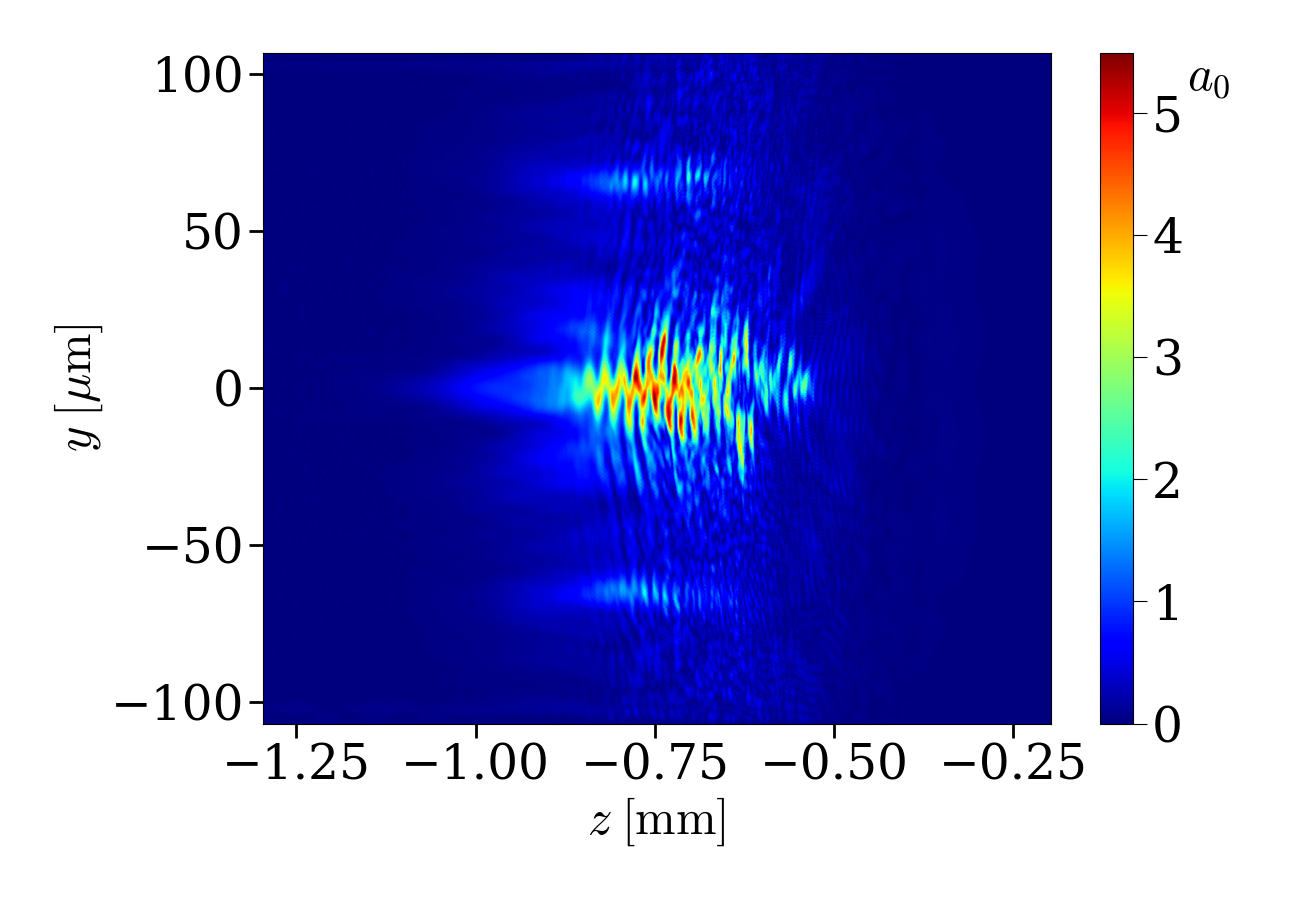}
    \caption{}
    \label{fig:sim3Dlas2}
\end{subfigure}
\begin{subfigure}{0.33\textwidth}
    \includegraphics[width=\textwidth]{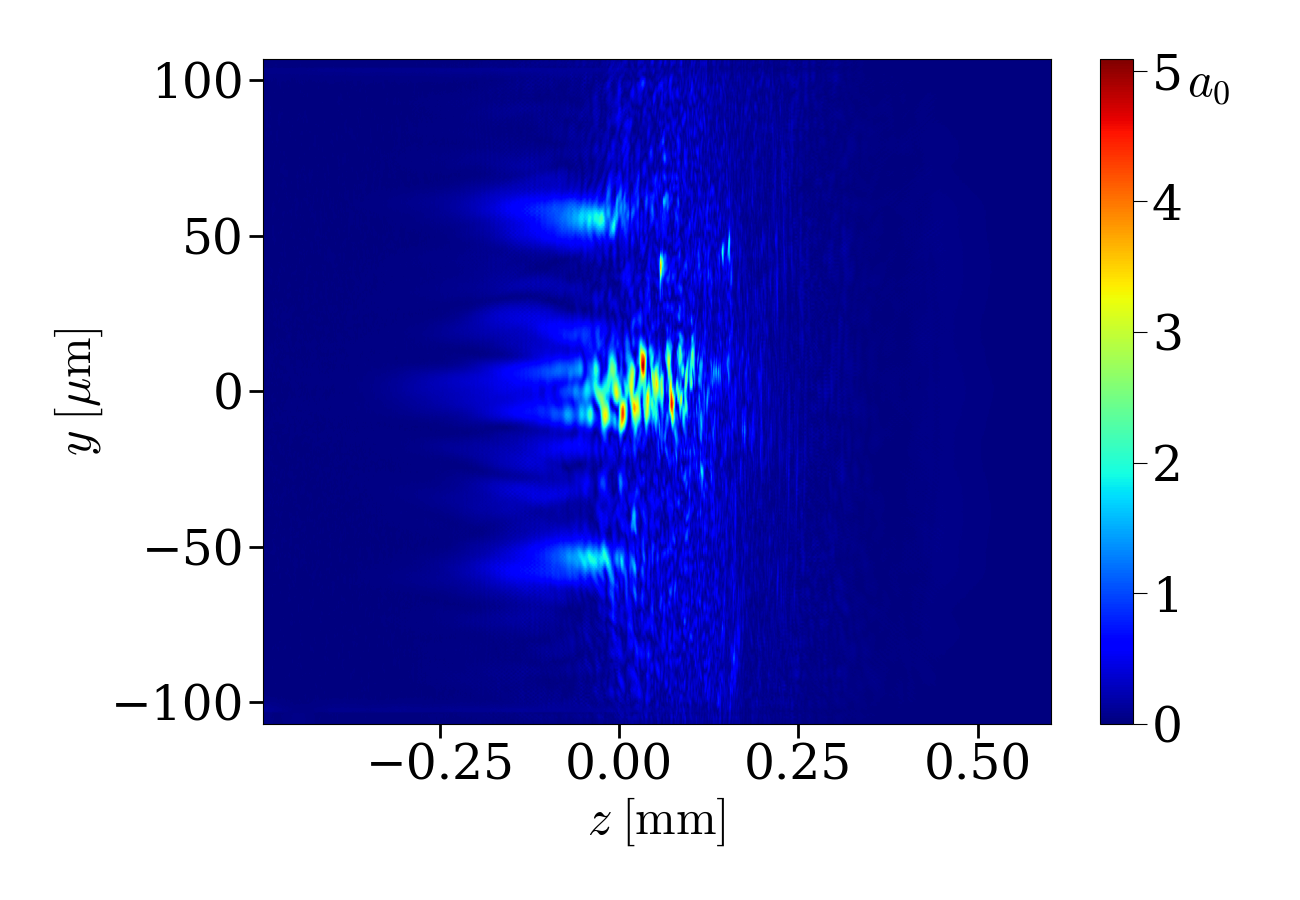}
    \caption{}
    \label{fig:sim3Dlas3}
\end{subfigure}
\caption{(a, c--e) 3D CALDER PIC simulation. (a) Evolution of the normalised peak amplitude ($a_0$) and energy ($E_{\rm laser}$) of the laser pulse as a function of propagation distance. (c--e) 2D cuts ($x=0$) of the normalised laser amplitude after three propagation distances: (c) $z \approx 0.1\,\rm mm$, (d) $z \approx 0.75\,\rm mm$ (d), and (e) $z \approx 1.15\,\rm mm$. (b) Quasi-3D OSIRIS simulation showing the distribution of energy gain induced by the wakefield ($W_{\rm LWFA}$) or DLA ($W_{\rm DLA}$) as a function of total electron energy $E$ for selected tracked particles; the dotted line represents $W=E$.}
\label{fig:sim3D}
\end{figure}

Figure~\ref{fig:sim3Da0} shows the evolution of $a_0$ and the laser energy with the propagation distance $z$. The simulation starts at $z = -3.5\,\rm mm$. Figure~\ref{fig:sim3Dlas1} shows the laser field envelope after 2.5~mm of propagation through the entrance density ramp (near $z = -1\,\rm mm$):
$a_0$ has then increased to $\sim 9$ due to relativistic self-focusing (corresponding to a intensity enhancement by a factor of $\sim 25$ up to $\sim 10^{20}\,\rm W\,cm^{-2}$) but self-modulation of the laser envelope remains weak. At later times, self-modulation becomes clearly visible (Figs.~\ref{fig:sim3Dlas2} and \ref{fig:sim3Dlas3}), and $a_0$ subsequently decreases due to diffraction and depletion. After 6~mm of propagation (Fig.~\ref{fig:sim3Da0}), most of the laser energy is depleted and $a_0$ is strongly reduced. This observation aligns well with experimental side-view images (Fig.~\ref{fig:setup}), where bright scattered light from the intense interaction is visible primarily in the first half of the gas jet.

The simulation also reveals (not shown) that the self-modulated laser beam drives a nonlinear wakefield, reaching a longitudinal accelerating electric field amplitude above the linear cold-plasma limit $m_e c\omega_p/e$, where $\omega_p$ is the plasma frequency. Electron acceleration occurs predominantly within the first 4~mm (up to $z \sim 0.5\,\rm mm$). Beyond this point, despite $a_0$ remaining around 3--4, the electron beam spectrum and charge cease to evolve due to the high energy already reached and the low remaining laser energy.

To clarify the acceleration mechanism and disentangle the respective contributions of wakefield acceleration and DLA, we also performed a quasi-3D PIC simulation using azimuthal Fourier mode decomposition~\cite{Lifschitz_JCP_2009}. This geometry enables us to isolate the energy gain contributions from the laser and plasma fields~\cite{King_PRL_2021, Miller_POP_2023}. From this simulation, performed using the OSIRIS code~\cite{fonseca2002, Davidson_JCP_2015} (see Methods for details), several electron trajectories were tracked. The energy gain induced by the axisymmetric longitudinal wakefield $E_z^{m=0}$ (where $m$ is the azimuthal mode index), $W_{\rm LWFA} = -e \int E_z^{m=0} v_z \, dt$, was compared directly to the gain induced by the laser field ($m = 1$), $W_{\rm DLA} = -e \int \mathbf{E}^{m=1} \cdot \mathbf{v} \, dt$, where $\mathbf{v}$ is the electron velocity. These two components are plotted against the total energy gain $E = W_{\rm LWFA} + W_{\rm DLA}$ in Fig.~\ref{fig:simOsiris}.
The results demonstrate that most of the charge is pre-accelerated to low or moderate energies by the wakefield. This confirms a feature observed in the 3D CALDER simulation: the wakefield phase velocity fluctuates rapidly due to dynamic evolution and self-modulation of the laser driver, causing accelerated electrons to quickly dephase into decelerating regions. However, as shown in Fig.~\ref{fig:simOsiris}, DLA provides a substantial boost to higher energies. Crucially, the high-energy tail of the electron spectrum is preferentially accelerated by DLA.

\subsection*{Hard X-ray beam}

\begin{figure}[htbp]
\centering
\begin{subfigure}{0.45\textwidth}
\includegraphics[width=\textwidth]{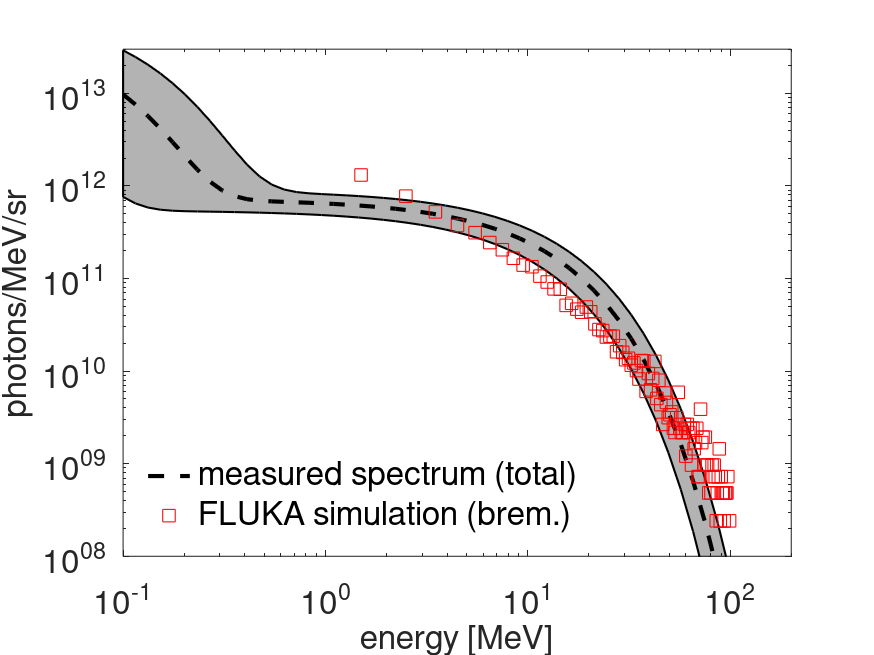}
\caption{}
\label{fig:Xa}
\end{subfigure}
\begin{subfigure}{0.5\textwidth}
\includegraphics[width=\textwidth]{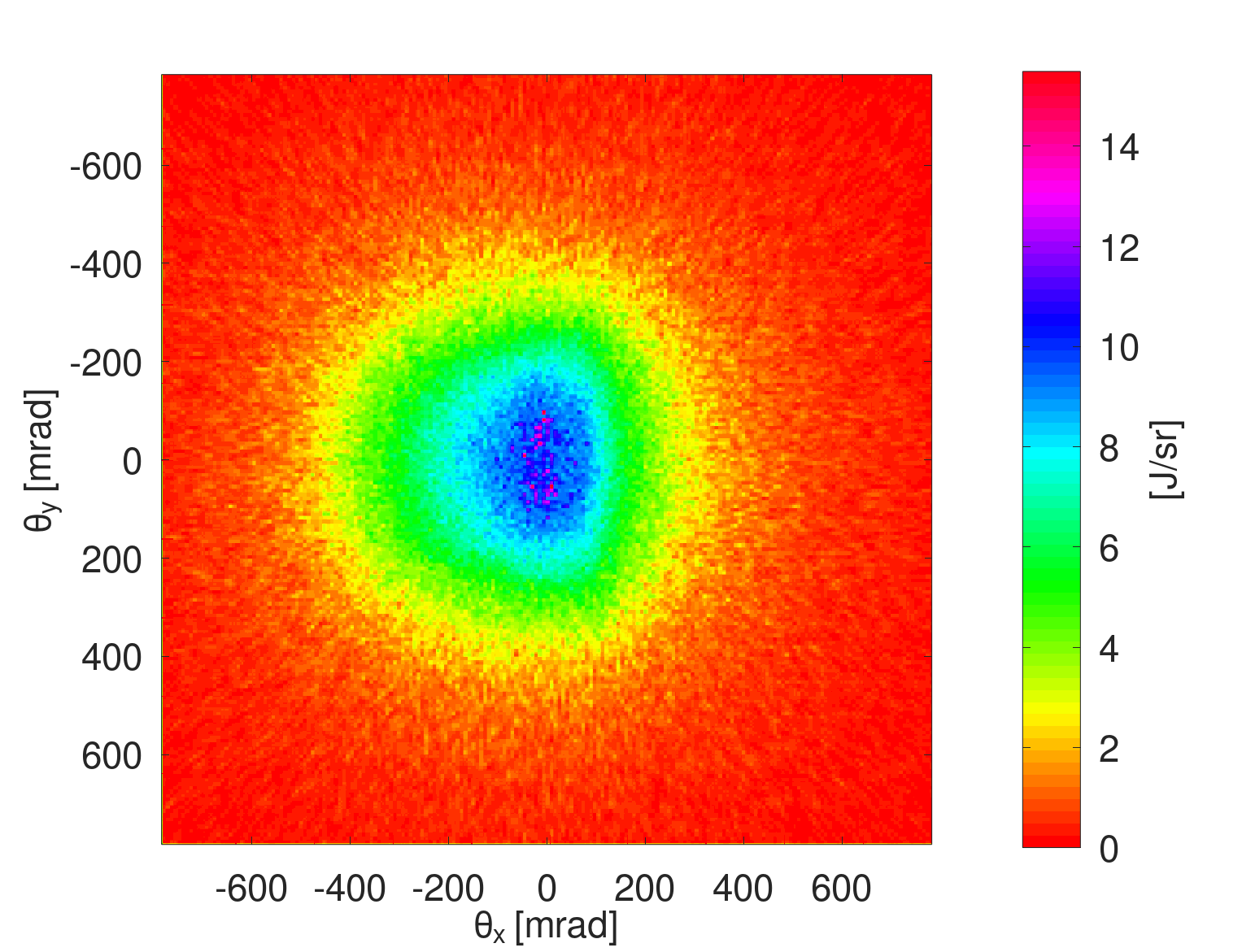} 
\caption{}
\label{fig:Xb}
\end{subfigure}
\caption{(a) Experimental X-ray spectrum retrieved from the cannon measurement for shot~\#5 (dashed line) and calculated Bremsstrahlung spectrum from FLUKA simulations (squares). The shaded area represents the confidence interval of the experimental fit. (b) Bremsstrahlung X-ray profile at a distance of 50~cm from the target chamber center (TCC), calculated from FLUKA simulations.}
\label{fig:X}
\end{figure}

Figure~\ref{fig:Xa} shows the X-ray spectrum retrieved from the cannon measurements, for shot~\#5. The data are best-fitted using a two-exponential spectrum $dN/dE \propto  e^{-E/T_c} + b\, e^{-E/T_h}$, with a ``cold'' temperature $T_c \simeq 55 \pm 33\,\rm keV$ and a ``hot'' temperature $T_h \simeq 9.3 \pm 1.1\,\rm MeV$, yielding a mean photon energy of $6.4 \pm 1.6\,\rm MeV$. Although betatron radiation \cite{rousse2004production} produced by electrons in the wakefield is a likely candidate for the emission in the tens-of-keV region, the resolution of our cannon spectrometer in the sub-$100\,\rm keV$ regime is insufficient to unambiguously identify its nature.

By contrast, the hot component clearly results from the expected Bremsstrahlung in the converter and the surrounding aluminum frame of the CRACC envelope. To confirm this, we carried out additional Monte Carlo simulations using the FLUKA code \cite{bohlen2014fluka, fasso2005fluka}. We implemented a simplified version of the CRACC geometry in the model and considered an electron beam with an energy spectrum and a transverse distribution representative of the experiment, though a homogeneous space-energy distribution was chosen for simplicity. The simulated Bremsstrahlung spectrum (red squares in Fig.~\ref{fig:Xa}) is in good agreement with the experimental curve in the high-energy region ($> 1\,\rm MeV$). Minor discrepancies may result from uncertainties in the measured electron beam spectrum, particularly regarding its space-energy distribution. 

This simulation also allows us to evaluate photon beam propagation in the LMJ chamber. An example of the transverse photon beam profile at 50~cm from TCC is shown in Fig.~\ref{fig:Xb}. The horizontal asymmetry of the beam arises from the decentered converter in our experiment (see Fig.~\ref{fig:setup}).

The energy carried by the photon beam can be inferred from its simulated transverse profile and measured spectrum, yielding a total hard X-ray energy of $E_{\rm X} = 3.3\,\rm J$ and an effective laser-to-X-ray energy conversion efficiency ($E_{\rm X}/E_{\rm use}$) of $1.7\%$. Importantly, the higher uncertainty in the cold component of the photon spectrum does not significantly impact these numbers. Over $99\%$ of the X-ray energy lies in the $>1\,\rm MeV$ region, given that the average photon energy is several MeV. While the converter in this experiment was designed specifically to study induced photonuclear reactions in the CRACC plates via $\gamma$-spectrometry (to be detailed in a separate publication), future optimization of the converter material and thickness could readily produce hard X-ray beams exceeding $10\,\rm J$.

In comparison, Günther \textit{et al.}~\cite{Gunther_NatComm_2022} reported a laser-to-photon conversion efficiency of $1.4\%$ for photon energies above $10\,\rm MeV$, corresponding to a yield of $10^{12}\,\rm photons/sr$. In our case, despite the non-optimized setup, the yield above $10\,\rm MeV$ is estimated to be approximately $2 \times 10^{12}\,\rm photons/sr$. Across the broader energy range above $1\,\rm MeV$, the photon flux exceeds $10^{13}\,\rm photons/sr$. Such a multi-MeV photon source opens new perspectives for applications, particularly for electron-positron pair creation and radioisotope production~\cite{Sun2021, Zhang2024, tavana2023ultra, tavana2026ultrahigh}.

\vspace{5mm}
In summary, we have demonstrated the efficient production of ultrahigh-charge, relativistic electron beams at the high-energy LMJ facility using the PETAL laser beam with an on-target power up to $0.95\,\rm PW$. We achieved charge values as high as $1.1 \pm 0.13\,\rm \mu C$ (above $0.9\,\rm MeV$), which, to our knowledge, represents the highest charge ever reported for a laser-wakefield accelerator. Above $10\,\rm MeV$, the estimated charge is $467\,\rm nC$, and above $100\,\rm MeV$, it reaches $17\,\rm nC$. PIC simulations reproduce the electron energy distribution, relying on well-monitored laser beam properties and gas density profiles. A mixed acceleration regime is revealed, where SMLWFA is mostly responsible for the high-charge trapping and pre-acceleration, while DLA significantly contributes to a high-energy boost.

The $1.1\,\rm \mu C$ electron beam carries a total energy of $\sim 17\,\rm J$. This energy was subsequently converted into more than $3\,\rm J$ of hard X-rays via Bremsstrahlung within a stack of metallic converters.The mean energies of the produced electron beams, on the order of a few tens of MeV, are well suited for applications in nuclear and laboratory astrophysics. The high charge density could also enable laboratory studies of instabilities produced by relativistic electron beams in auxiliary plasma targets \cite{bret2010multidimensional, PhysRevResearch.4.023085, PhysRevResearch.3.023103, Arrowsmith2025SuppressionOP}. Secondary photons generated by Bremsstrahlung--or potentially by other processes such as betatron oscillations or inverse Compton scattering \cite{ta2012all}--can also be used in combination with powerful LMJ beams for probing matter in HED conditions \cite{albert2016applications}. These possibilities will be further explored in the coming years at the LMJ-PETAL facility.

\section*{Methods}

\subsection*{CRACC imaging plate analysis}

\subsubsection*{Overall process}

The IP signal is initially given in units of photostimulated luminescence (PSL), which corresponds to the effective signal obtained after scanning. Due to the high fluence of electrons impacting the IP detector, the readout signal after a single scan was heavily saturated. IPs thus had to be scanned repeatedly ($N > 30$) until saturation was eliminated, which, however, caused an overall attenuation of the PSL level. The effective PSL signal used to recover the charge is therefore the PSL signal measured after the $N^{\mathrm{th}}$ scan multiplied by the cumulative attenuation factor from multiple scans.

The attenuation was measured in region(s) where pixels remained unsaturated during the first scan. The attenuation factor is then simply given by the ratio $\langle \mathrm{PSL}_{1^\mathrm{st}\, \mathrm{scan}} \rangle /\langle \mathrm{PSL}_{N^\mathrm{th}\,\mathrm{scan}}\rangle$ in this region. In cases where all pixels were saturated even during the initial scan(s), we relied on typical attenuation curves measured from other scan series. Notably, our method differs from that used in Ref.~\cite{Sha21}, where the attenuation factor was inferred from an exponential fit of the decreasing PSL signal as a function of scan number. Applying that approach to our dataset would have significantly overestimated the attenuation factor--and consequently the final charge and conversion efficiency values--by a factor of $\sim 1.5$--10, depending mainly on the total number of scans. Signal loss due to fading was also accounted for, as IPs were typically scanned $\sim 1000\,\rm min$ after the shot~\cite{ohuchi2000functional, Bou15}.

Because a hole was drilled through the IPs to allow the on-axis portion of the electron beam to propagate to the SESAME spectrometer, the central signal on the IPs is missing. To reconstruct this region, a least-squares fit of the measured IP signal was performed using a sum of two-dimensional Gaussian functions, which was then extrapolated across the missing central area. Finally, the total charge is evaluated over an elliptical area whose semi-axes equal twice the standard deviation of the horizontal and vertical lineouts of the fitted beam profile (corresponding to the second-moment beam width definition).

\begin{figure}[htbp]
\centering
\begin{subfigure}{0.3\textwidth}
\includegraphics[width=\textwidth]{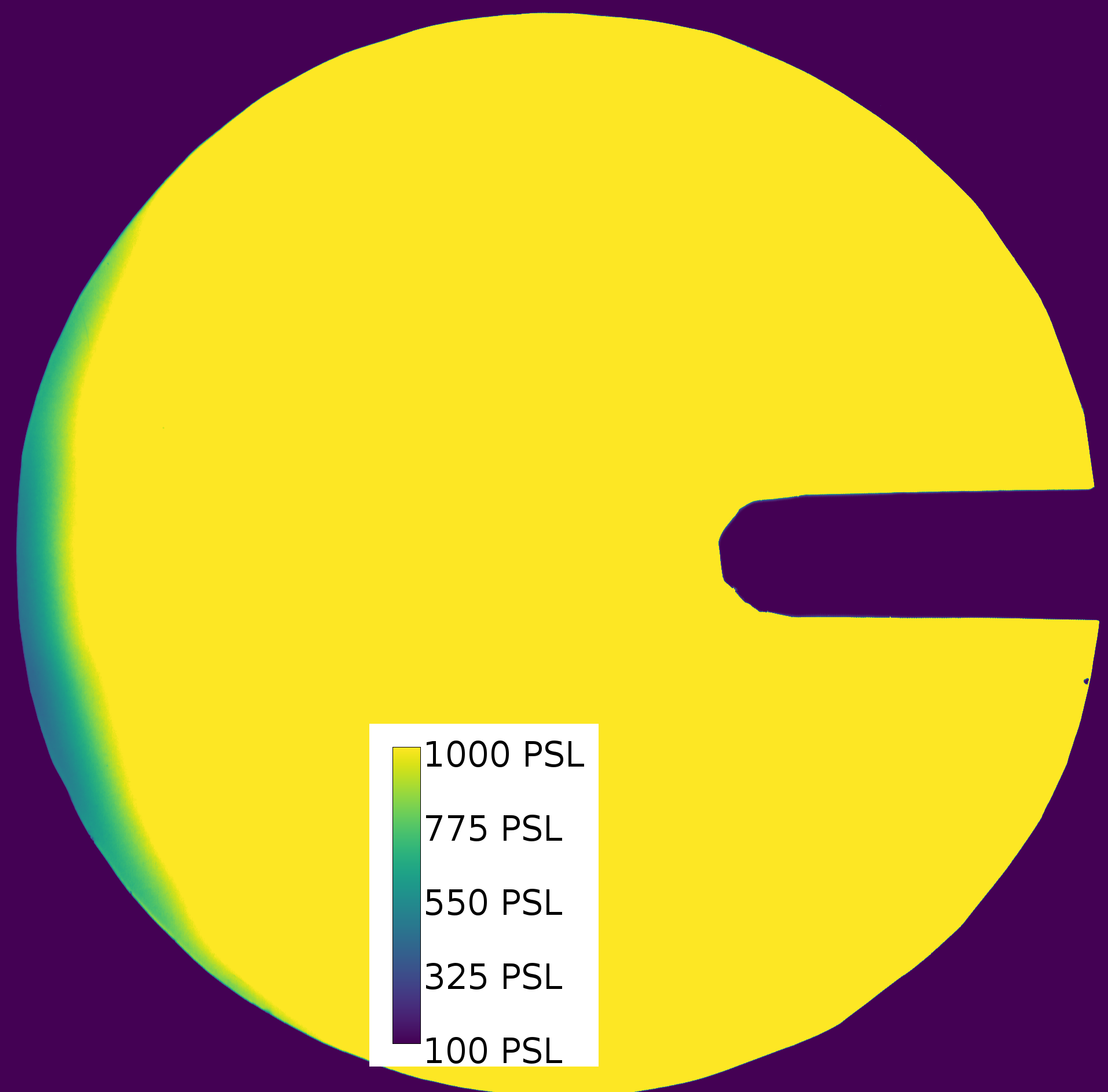}
\caption{}
\label{fig:IPscan1}
\end{subfigure}
\begin{subfigure}{0.3\textwidth}
\includegraphics[width=\textwidth]{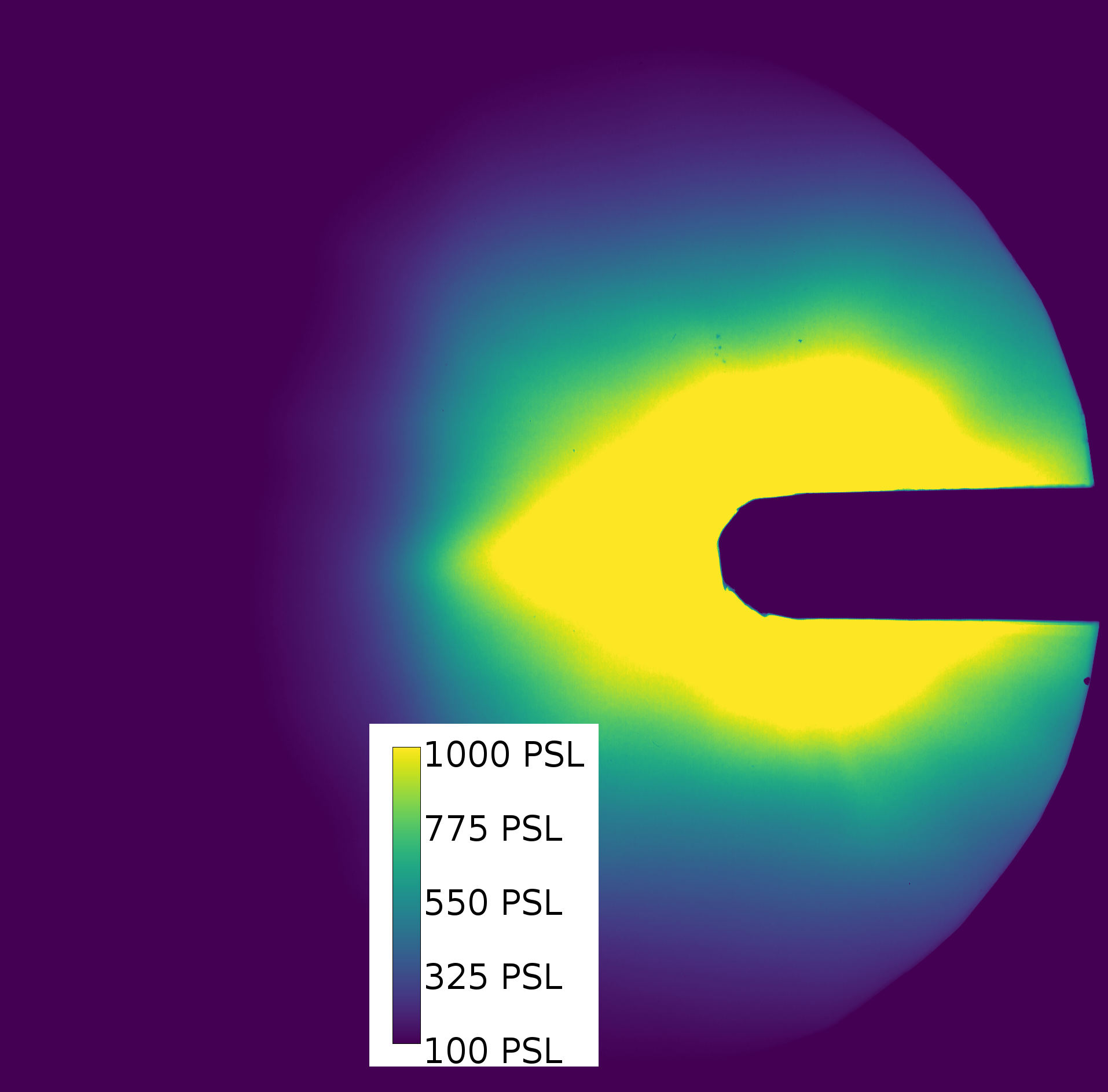} 
\caption{}
\label{fig:IPscan15}
\end{subfigure}
\begin{subfigure}{0.3\textwidth}
\includegraphics[width=\textwidth]{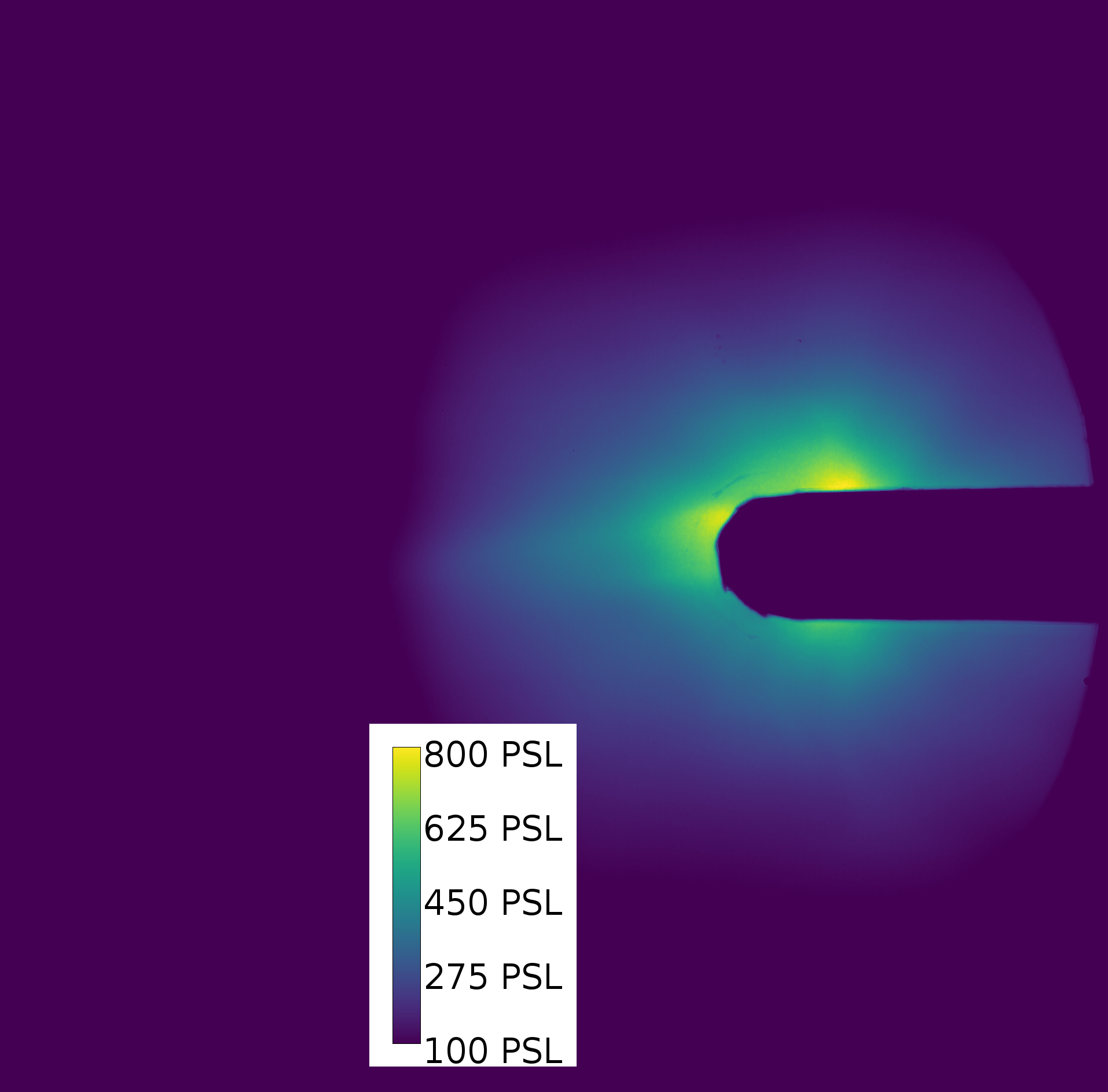} 
\caption{}
\label{fig:IPscan30}
\end{subfigure}
\caption{Example of PSL signal distribution obtained after 1 scan (a), 15 scans (b) and 30 scans (c) of the CRACC IP.}
\label{fig:IPscan}
\end{figure}

\subsubsection*{Uncertainty}

The main source of uncertainty is associated with the calculation of the attenuation factor. This calculation can be performed by tracking a single region of interest (ROI) across all IP scans. However, using a single ROI yields a final signal level $\langle \mathrm{PSL}_{N^{\mathrm{th}}\,\mathrm{scan}} \rangle$ that is very low and close to the noise floor. Because the saturated area shrinks during the scanning process, it is possible to dynamically update the ROI after a few scans to maintain a sufficiently high PSL signal level in the selected area. In this approach, the total attenuation factor is given by the product of the intermediate attenuation factors across successive ``fresh'' ROIs, $\left.\frac{\langle \mathrm{PSL}_{1^\mathrm{st}\,\mathrm{scan}} \rangle}{\langle \mathrm{PSL}_{j^\mathrm{th}\,\mathrm{scan}} \rangle } \right|_{1^\mathrm{st}\,\mathrm{ROI}} \times \left.\frac{\langle \mathrm{PSL}_{j^\mathrm{th}\, \mathrm{scan}} \rangle}{\langle \mathrm{PSL}_{k^\mathrm{th}\,\mathrm{scan}} \rangle} \right|_{2^\mathrm{nd}\,\mathrm{ROI}} \times ... \times \left.\frac{\langle \mathrm{PSL}_{m^\mathrm{th}\,\mathrm{scan}} \rangle}{\langle \mathrm{PSL}_{N^\mathrm{th}\,\mathrm{scan}} \rangle} \right|_{\rm last\,ROI}$. 
These two approaches--single ROI and multiple ROIs--serve to define a lower bound and an upper bound, respectively, for the attenuation factor, and consequently for the reconstructed beam charge.

The second main source of uncertainty is related to the estimation of the signal in the aperture region of the IPs. To account for this, we include an additional $10\%$ uncertainty in the reported charge, based on typical variations in the fitting parameters.

\subsubsection*{IP response}

The response function of the IP (in units of PSL per incident electron) was calculated via Geant4 simulations for monoenergetic electrons ranging from $100\,\mathrm{keV}$ up to the maximum energy measured by SESAME. The exact CRACC geometry, layer stack, and materials were imported into the simulations. These therefore account for both the direct electron beam signal and noise contributions (such as backscattered electrons or Bremsstrahlung generated in the CRACC frame) that deposit non-negligible energy into the active IP layers. In this way, the IP response is calculated with high accuracy, preventing an overestimation of the incoming beam charge.

\begin{figure}[htbp]
\centering
\includegraphics[width=.8\textwidth]{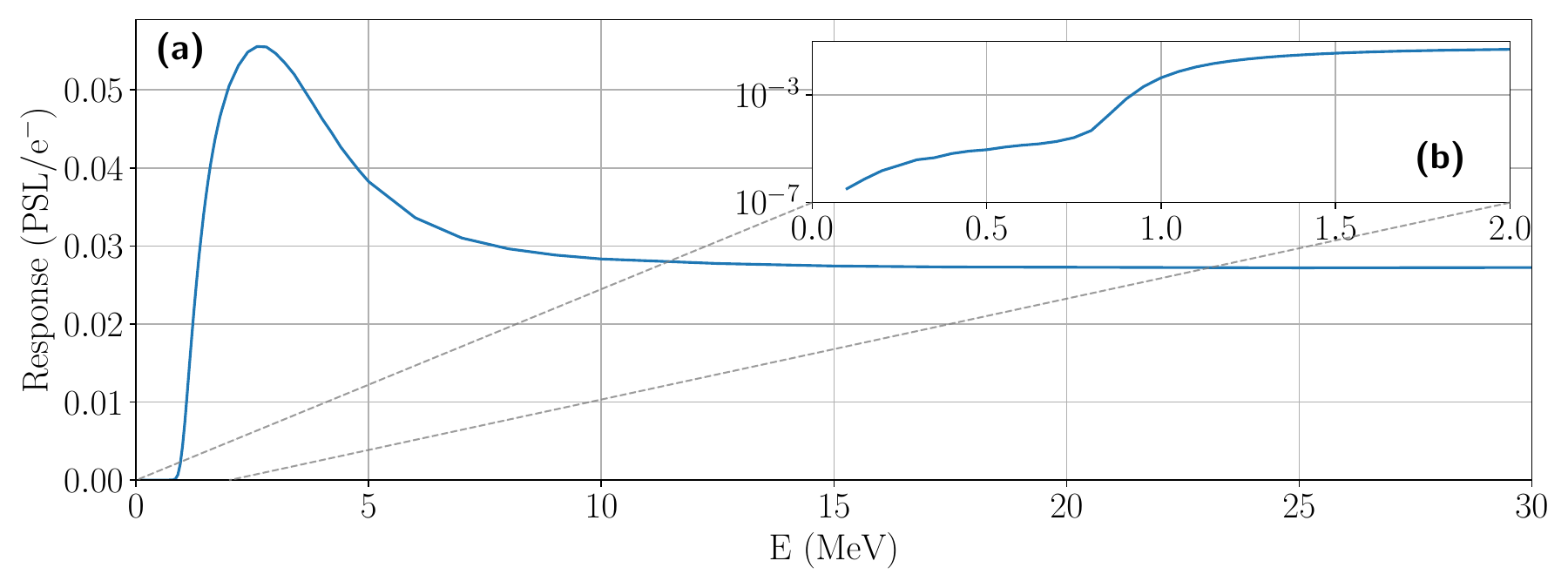} 
\caption{Example of calculated IP response for one of the shots, in (a) linear scale and (b) logarithmic scale with an emphasis on the low-energy region.}
\label{fig:IPresponse}
\end{figure}

An example of the calculated IP response is shown in Fig.~\ref{fig:IPresponse}. The IP response below $\sim 0.8$--0.9~MeV is very weak ($< 0.1\,\mathrm{mPSL/e^-}$), meaning that the reported charge values effectively represent electrons above this energy threshold. Indeed, the IP is preceded by a millimeter-thick CRACC cover (high-density polyethylene), aluminized Mylar foils to block laser light, and, on certain shots, an additional gold foil. Together, these layers act as a filter against low-energy electrons before they can reach the IP. For each shot, the IP response function was weighted using the energy spectrum measured by SESAME. Finally, the total integrated PSL signal was converted into total electron yield, providing the beam charge.

\subsection*{Electron spectrum measurement and analysis}

\begin{figure}[htbp]
\centering
\includegraphics[width=1\textwidth,trim={1cm 3cm 1cm 3cm},clip]{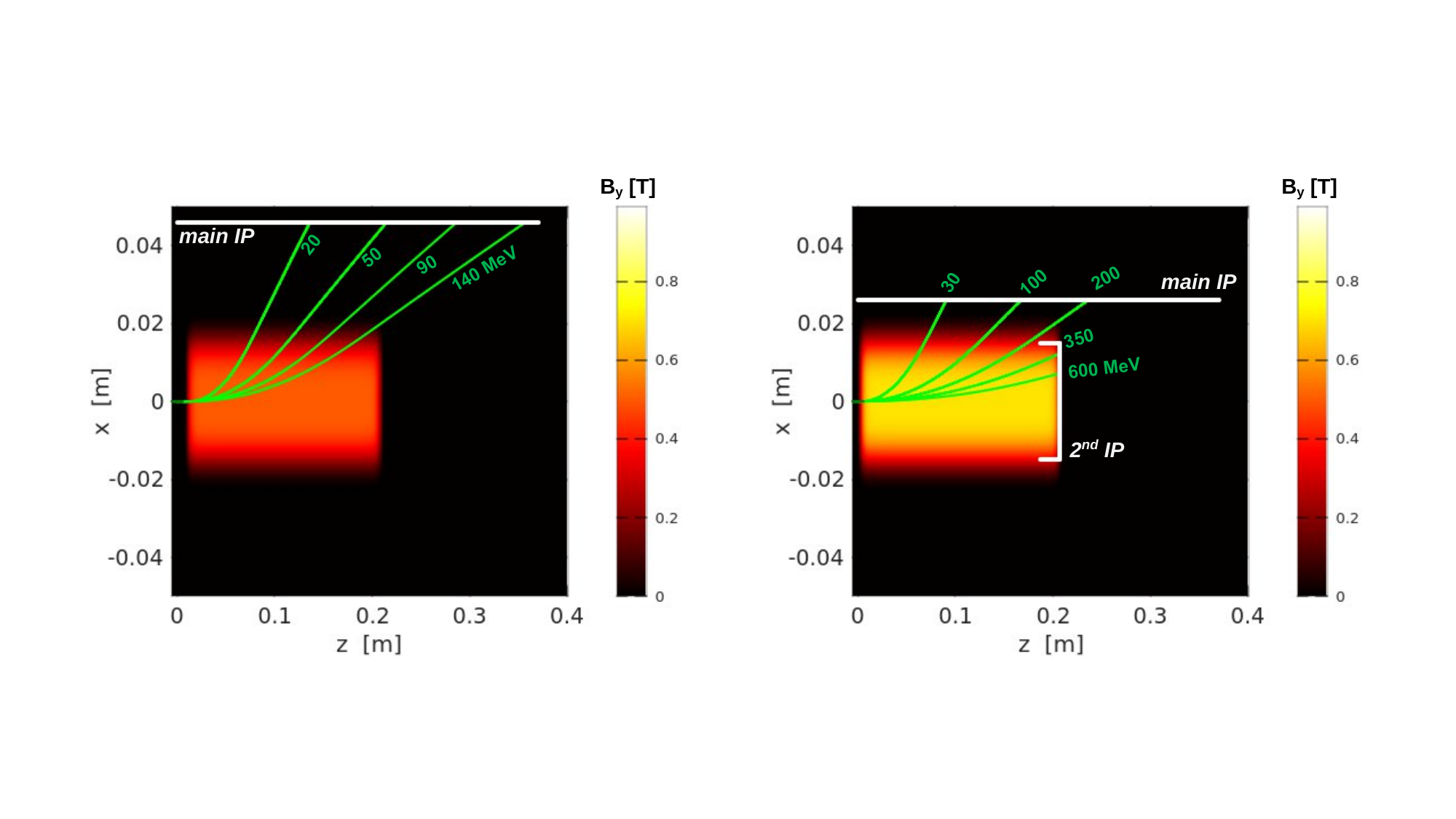} 
\caption{Calculated trajectories (green plain lines) for on-axis electrons in SESAME, as seen from a top view of the detector. The electrons propagate from the left to the right. Left: configuration corresponding to shots~\#1--4, with a maximum magnetic field $\sim$0.5~T (trajectories for electron energies of 20, 50, 90 and 140 MeV are represented). Right: configuration corresponding to shot~\#5, with a maximum magnetic field $\sim$0.7~T (trajectories for electron energies of 30, 100, 200, 350 and 600 MeV are represented). The positions of the IPs are represented with the white plain lines. The colorbar represents the magnitude of the perpendicular field component $B_y(x,z)$, as measured with a Hall probe scan.}
\label{fig:SESAME}
\end{figure}

SESAME is primarily composed of a $5\,\rm mm$ entrance slit and a permanent dipole magnet that deflects electrons onto a lateral image plate (IP). For shots~\#1--4, a $0.5\,\rm T$ magnet was used, and the lateral IP was placed at a distance of 46~mm from the optical axis, limiting the maximum detected energy to $\sim 150\,\rm MeV$. For shot~\#5, SESAME was upgraded by using a 0.7~T magnet, positioning the lateral IP closer (26~mm from the optical axis), and adding a second IP at the end of the magnet to collect the highest-energy electrons. The dispersion functions were obtained from calculated electron trajectories in both configurations, as shown in Fig.~\ref{fig:SESAME}. In the upgraded configuration, the additional IP was folded to accommodate mechanical constraints. It was thus composed of a lower-energy region (240--280~MeV) parallel to the main IP, and a higher-energy region ($> 340\,\rm MeV$) perpendicular to the main IP. The fold in the second IP degraded the signal in the intermediate 280--340~MeV range, which is therefore omitted from Fig.~\ref{fig:spectra}. Both dipole magnets were characterized using Hall probe measurements along with complementary calibration performed on a linear accelerator beamline.

The measured electron spectrum $d^2N/(dE\,d\Omega)$ was fitted with a two-temperature Maxwellian distribution, denoted $S(E)$. We then evaluated the weighted mean energy of the on-axis spectrum, given by $\langle E \rangle = \frac{\int E\, S(E)\, dE}{\int S(E)\, dE}$. The weighted mean energy was calculated from the fitted distribution $S(E)$ rather than directly from the raw signal measured by SESAME. This approach better accounts for low-energy electrons not detected by SESAME, avoiding an overestimation of the mean energy value. Integration was performed between 0.9~MeV (corresponding to the lowest electron energy considered for the reported charge values) and the maximum energy measured by SESAME (150~MeV for shots~\#1--4 and 550~MeV for shot~\#5).

\subsection*{Gas density}

\begin{figure}[htbp]
\centering
\includegraphics[width=.6\textwidth]{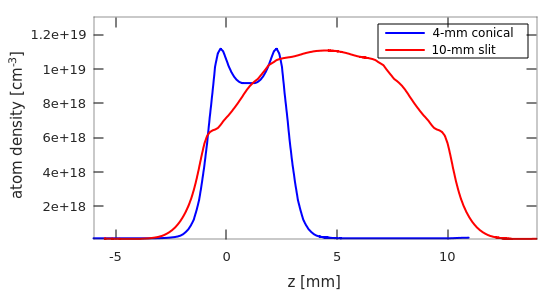} 
\caption{Neutral gas density profiles obtained for helium at a backing pressure of 50 bars along the optical axis, at a distance of 3~mm above the nozzle. Profiles were calculated with CFD simulations. The geometric focus of PETAL was set at the longitudinal position $z=0$~mm inside the jet.}
\label{fig:gasdensity}
\end{figure}

Gas density profiles were calculated through computational fluid dynamics (CFD) simulations. We relied on the OpenFOAM code using the \textit{HiSA} solver with Reynolds-averaged Navier-Stokes $k-\omega$ SST model\cite{openfoam,hisa}.  For better accuracy, peak density values and shot-to-shot variations were also characterized experimentally by means of interferometric measurements performed on a dedicated experimental bench that replicates the LMJ gas installation system.
The simulated profiles were finally normalized to the experimentally inferred peak values, with an associated uncertainty of the order of $10\%$.
The corresponding gas density profiles, that were then injected in the PIC simulations, are shown in Fig.~\ref{fig:gasdensity}.

\subsection*{Particle-in-cell simulations}

In the 3D CALDER simulations (shots~\#4 and \#5), the simulation box length and transverse size are $1.1\,\rm mm$ and $214\,\rm \mu m$, respectively. A moving window is used to follow the laser propagation. The cell sizes in the longitudinal and transverse directions are $\Delta z = 0.5~c/\omega_0$ and $\Delta x = \Delta y = 4~c/\omega_0$, respectively. The time step is $\Delta t = 0.492\,\omega_0^{-1}$. Absorbing boundary conditions are applied on the transverse sides of the box. The B-TIS3 interpolation scheme \cite{Bourgeois_JPP_2023} is used to improve the numerical accuracy of the electron transverse motion and acceleration. One macro-particle per cell is used to describe the ion background, and four are used for the pre-ionized electrons.

In the 3D CALDER simulation of shot~\#5, the initial laser intensity is $2 \times 10^{19}\,\rm W\,cm^{-2}$, corresponding to a normalised peak vector potential $a_0 = 3.82$. The laser spot is approximated as the coherent sum of two Gaussian beams. The first beam is modeled by $a_1 = 2.97$, a focal spot size $w_{y1} = 23.4\,\rm \mu m$ and $w_{x1} = 34\,\rm \mu m$. The second beam is characterized by $a_2 = 0.85$, $w_{y2} = 260.5\,\rm \mu m$, $w_{x2} = 180.9\,\rm \mu m$. The laser beam duration is 753~fs. The plasma density reproduces the 10-mm nozzle density profile shown in Fig.~\ref{fig:gasdensity}.

In the quasi-3D simulation conducted with the OSIRIS code \cite{Davidson_JCP_2015}, only the first 4~mm (from $z=-3.5\,\rm mm$ to $z=0.5\,\rm mm$) of the interaction are simulated, as the electron spectrum does not evolve beyond this distance. The increasing plasma density in this range is modeled using the first half of a Gaussian fit. The axisymmetric (imposed by the quasi-3D geometry) laser spot is characterized by $a_0 = 1.78$ and $w_{x,y} = 30\,\rm \mu m$. Only the central hotspot is described, neglecting the laser energy in the focal spot wings. The temporal profile of the field amplitude is defined as $a_0 = \sin \left(\frac{\pi t}{2 \tau_{\rm las}}\right)^2$, where $\tau_{\rm las} = 682\,\rm fs$. Only the first two azimuthal modes ($m = 0, 1$) are used. A moving window is employed, with dimensions of $L_z \times L_r = 540 \times 86.5\,\rm \mu m$ ($L_r$ corresponds to the radius of the box), using 60 cells per laser wavelength longitudinally and 20 cells transversely. A dual electromagnetic solver is used~\cite{Li_CPC_2021}, with a macro-particle resolution of $N_x \times N_r \times N_\theta = 2 \times 1 \times 8$ per cell.

Due to the simplified description of the laser and plasma profiles, combined with the reduced quasi-3D geometry's inability to capture non-axisymmetric laser envelopes (which can be observed after self-modulation in Fig.~\ref{fig:sim3D}), this quasi-3D OSIRIS simulation is intended only to qualitatively reproduce the physics. Nevertheless, key trends similar to those in the 3D CALDER simulation are recovered: the laser beam initially self-focuses, subsequently self-modulates, and ultimately depletes. The electron acceleration length, spectrum evolution, and maximum achievable energy are also in close agreement.

\bibliography{bib}

\section*{Acknowledgements}

The PETAL project has been performed by the CEA (“maître d’oeuvre”) under the financial auspices of the New Aquitaine Region in France (“maître d’ouvrage”, project owner), the French Government, and the European Union. This work is supported by the New Aquitaine Region through the PETAL-UPgrade project (contracts \#13532820,  \#22577420 and \#42979220).
The CRACC and SESAME diagnostics were designed and commissioned as a result of the PETAL+ project coordinated by the University of Bordeaux and funded by the French Agence Nationale de la Recherche under grant ANR-10-EQPX-42-01. LMJ-PETAL experiments presented in this article were supported by the Association Lasers et Plasmas and by CEA.
J.C.K. is acknowledging the support of Natural Sciences and Engineering Council of Canada.
We acknowledge the Grand Equipement National de Calcul Intensif GENCI-TGCC for providing us access to the supercomputer IRENE under Grant No. A0190512993.

\section*{Author contributions statement}

F.Al., D.B, J.L.B, X.D., J.C.K., N.L. and K.T.P. proposed the initial experiment.\\
G.B., M.B, T.C., A.D., R.D.J, S.D, W.D., R.D., B.E., M.F., J.G., V.H., E.J., E.L, I.L., L.L.D., R.P., C.R. and B.V. developed the diagnostics, targetry and operated the experiment.\\
N.B., F.Au., C.C., S.C., F.S. and W.V. developed and operated the laser.\\
W.C. and B.M. coordinated the experiments.\\
G.B., B.M. and L.R. analyzed the experimental data.\\
F.Al., G.B., W.C., X.D., E.D.H, L.G., I.L., B.M., L.R. and B.V. discussed the results.\\
R.B., P.C., X.D., M.G., B.M., L.R. and M.V. performed the simulations.\\
W.C., X.D., L.G., B.M. and L.R. wrote the first version of the manuscript.\\
All authors reviewed the manuscript.\\

\section*{Additional information}

The authors declare no competing interests.

\end{document}